\documentclass[12pt]{article}\usepackage[hyperfootnotes=false]{hyperref}
\usepackage{epsfig}
\usepackage{float}
\usepackage{dsfont}

\usepackage{bbold}

\usepackage{caption}
\usepackage{subcaption}

\usepackage{amsmath}
\usepackage{amssymb}
\usepackage{graphicx}
\newlength{\abstractwidth}
\renewcommand{\thefootnote}{\fnsymbol{footnote}}
\renewcommand{\thanks}[1]{\footnote{#1}} 
\newcommand{\starttext}{
\setcounter{footnote}{0}
\renewcommand{\thefootnote}{\arabic{footnote}}}

\newcommand{\be}{\begin{equation}}
\newcommand{\bea}{\begin{eqnarray}}
\newcommand{\eea}{\end{eqnarray}}
\newcommand{\beq}{\begin{equation}}
\newcommand{\ee}{\end{equation}}

\def\eq{&=&}
\def\d{\partial}

\def\la{\langle}
\def\ra{\rangle}

\def\simleq{\; \raise0.3ex\hbox{$<$\kern-0.75em
\raise-1.1ex\hbox{$\sim$}}\; }
\def\simgeq{\; \raise0.3ex\hbox{$>$\kern-0.75em
\raise-1.1ex\hbox{$\sim$}}\; }

\def\bi{\begin{itemize}}
\def\ei{\end{itemize}}

\def\sc{\setcounter{equation}{0}}

\def\dof{degrees of freedom }

\def\CA{{\cal{A}}}

\def\CC{{\cal{C}}}

\def\CO{{\cal{O}}}

\def\CT{{\cal{T}}}
\def\CU{{\cal{U}}}

\def\CA{{\cal{A}}}

\def\qt{quantum teleportation \ }

\def\bn{\bigskip \noindent}

\makeatletter
\g@addto@macro\normalsize{%
  \setlength\abovedisplayskip{10pt}
  \setlength\belowdisplayskip{20pt}
  \setlength\abovedisplayshortskip{10pt}
  \setlength\belowdisplayshortskip{20pt}
}
\makeatother

\usepackage{color}

\renewcommand{\title}[1]{\vbox{\center\LARGE{#1}}\vspace{5mm}}
\renewcommand{\author}[1]{\vbox{\center#1}\vspace{5mm}}
\newcommand{\address}[1]{\vbox{\center\em#1}}

\begin{document}
  
\begin{titlepage}

\rightline{}
\bigskip
\bigskip\bigskip\bigskip\bigskip
\bigskip

\centerline{\Large \bf {Teleportation Through the Wormhole}}

\bn

\bigskip

\bigskip
\begin{center}

\author{Leonard Susskind and Ying Zhao}

\address{Stanford Institute for Theoretical Physics and Department of Physics, \\
Stanford University, Stanford, CA 94305-4060, USA}

\end{center}

\begin{center}
\bf     \rm

\bigskip

\end{center}

\begin{abstract}

ER=EPR allows us to think of quantum teleportation as
 communication of  quantum
information through space-time wormholes connecting entangled systems. The conditions for teleportation 
 render the wormhole traversable so that a quantum system entering one end of the ERB will, after a suitable  time, appear at the other end. Teleportation requires the transfer of classical information outside the horizon, but the classical bit-string carries no information about the  teleported system; the teleported system  passes through the ERB leaving no trace outside the horizon.  In general the teleported system will retain a memory of what it encountered in the wormhole. This  phenomenon could be observable in a laboratory equipped with quantum computers.

\medskip
\noindent
\end{abstract}

\end{titlepage}

\starttext \baselineskip=17.63pt \setcounter{footnote}{0}

\vfill\eject

\tableofcontents

\vfill\eject

\section{What is Quantum Teleportation and Why is it \\ Interesting?}

Quantum gravity is nothing if not surprising:

\bi 
\item Black holes are not black: they have entropy, temperature, and they evaporate.
\item Geometric theorems like the non-decrease of horizon area are violated.
\item The most fundamental locality  principle of quantum field theory---that degrees of freedom can independently be varied  in different regions of space---is not even approximately correct\footnote{We have in mind the holographic principle.}.
\item Quantum entanglement is responsible for the continuity of space \cite{VanRaamsdonk:2010pw}. 
   \ei
  
  \bn 

  These things have been around for some time and we have gotten used to them; they've become ``part of the furniture," so to speak\footnote{LS first heard it described this way by Geoff Pennington. }. Nevertheless they  were very unexpected.
  
The most recent surprise is the ER=EPR principle \cite{Maldacena:2013xja} that equates the existence of wormholes with quantum entanglement. One manifestation of ER=EPR is that under certain special conditions, Alice and Bob can jump into very distant black holes, and quickly meet  behind the horizon (see for example \cite{Marolf:2012xe}). This of course  is very interesting, but unfortunately it cannot be  observed  from outside the horizon. One might be led to believe that there is no operational meaning to ER=EPR, at least for observers outside the horizon.
This paper is about another quite distinct manifestation of ER=EPR---one which can be observed  from outside the horizon---``Teleportation through the wormhole" \cite{Susskind:2014yaa}. As explained  in \cite{Gao:2016bin}\cite{Maldacena:2017axo} it violates  another  classical property of GR, namely the non-traversability of wormholes   \\

Quantum teleportation was discovered about twenty five years ago by  C. H. Bennett, G. Brassard, C. Crépeau, R. Jozsa, A. Peres, W. K. Wootters, in a remarkable paper \cite{Bennett} ``Teleporting an Unknown Quantum State via Dual Classical and Einstein-Podolsky-Rosen Channels." The only new thing about teleportation through the wormhole, is that Einstein-Podolsky-Rosen equals Einstein-Rosen.

To illustrate just how remarkable quantum  teleporation is, let's compare it with a classical protocol to send a secure message, say of one bit. The classical protocol begins with Charlie holding two bits, $\bf A$ and $\bf B$. Either both bits are $0,$ or they are both $1.$ (The choice may have been made by consulting a pseudo-random number generator.) Charlie hands $\bf A$  to Alice and $\bf B$ to Bob who then carry them to some large relative separation. The probability is $1/2$ that $\bf A = \bf B \rm = 0,$  and $1/2$ that $\bf A = \bf B \rm = 1.$  

Alice also has another bit  $\bf T$---the teleportee. She wants to send $\bf T$ to Bob, so she  looks at both bits in her possession. If they are the same she sends Bob a message saying \it same. \rm Likewise if they are different she sends the message \it different\rm.  When Bob receives the message he either  flips $\bf B$ if the message says \it different \rm, or doesn't flip $\bf B$ if the message says \it same. \rm  In either case Bob's bit winds up in the same configuration as the bit $\bf T$. In effect Alice has sent 
  $\bf T$  to Bob. However, unlike the quantum case,  this classical protocol allows Alice to retain her own copy of $\bf T.$
  
  Now suppose Eve intercepts  Alice's message. What does she find out about $\bf{T}?$ Nothing, because she doesn't know the bit-values that Charlie handed off to Alice and Bob. 
  But is the protocol really perfectly secure? Charlie may have the memory (of which bit-values he handed off to Alice and Bob) stored in his brain. If so Eve could very gently probe Charlie. In classical physics she can probe Charlie so gently that he wouldn't feel it. Then, if she intercepted Alice's message, she could determinet the original configuration of $\bf T.$ Furthermore all of this could be done without disturbing the protocol.

  What if Charlie's memory was erased? In that case the bit of information will be emitted into the environment and Eve can in principle detect it. In fact classically many copies of Charlie's memory may be found, some still in Charlie's head, and others in the environment. Even classical gravitational radiation, from whatever manipulation Charlie did when  he handed Alice and Bob their bits, will store a perfect record that Eve can access\footnote{Remember that in classical physics an arbitrarily weak signal can carry arbitrary amounts  of information. }.
  The conclusion is that no  protocol, in a completely classical world,  can be perfectly secure against an eavesdropper with sufficient power.

\bn

  Now let us consider the quantum teleportation of a qubit. This time Charlie starts with two qubits in the entangled state $$|0\ra_A |0\ra_B + |1\ra_A|1\ra_B$$ (ignoring normalization). He hands one qubit to Alice and the other to Bob. Alice  has a third qubit $\bf T$ in the quantum state 
  $$|\Phi\ra_T \equiv \Phi(0)|0\ra_T + \Phi(1) |1\ra_T.$$  
  
 Alice measures her two qubit system---call 
  it $\bf AT$---in the Bell basis  
\bea
|1\ra \eq |00\ra + |11\ra   \cr  
|x\ra \eq |10\ra + |01\rangle \cr 
|y\ra \eq |10\ra - |01\ra   \cr  
|z\ra \eq |00\ra - |11\ra   
\eea
and gets one of four outcomes  labeled $(1,x,y,z).$ 
She then writes the outcome on a scrap of paper and sends it to Bob. When Bob gets the classical message, depending on what it says, he  applies one of four operators $1$, $X$, $Y,$ or $Z$ to his qubit. The result is that Bob's qubit  always ends up in the state 
  $$|\Phi\ra_B \equiv \Phi(0)|0\ra_B + \Phi(1) |1\ra_B,$$
   i.e. the original state of the  qubit  $\bf  T$. 
  
Can Eve successfully intercept the message and determine anything about the original state $|\Phi\ra_T$ of $\bf T$? The answer is no---in principle she cannot. The reason is the monogamy of entanglement; if the qubits $\bf A$ and $\bf B$ are maximally entangled, they cannot be correlated with any other system such as Bob's brain, the environment, or gravitational radiation. Therefore, unlike the classical case, quantum mechanics does not permit  Eve to learn anything about the state of $\bf T$.

A quantum-naive person might ask how the  information was transferred from Alice to Bob if it didn't pass through the space between them? The quantum-savvy person would answer that information in quantum mechanics is non-local. It does not make sense to ask where it is at any given time; it is non-locally distributed in the entangled state.

But that's not the final answer. If ER=EPR is to be believed, the qubit $\bf T$ was teleported through the microscopic Einstein-Rosen bridge, or wormhole, connecting the entangled  pair  shared by  Alice and Bob.  Of course for such a small system---a single Bell pair---the wormhole is too small to have a classical geometry, but we can apply the same reasoning  when the entangled system is a pair of macroscopic black holes \cite{Susskind:2014yaa}.  Then we may hope to follow the geometry of the wormhole as the teleported system passes through it \cite{Gao:2016bin}\cite{Maldacena:2017axo}.

Combining quantum teleportation with the idea that entangled black holes are connected by Einstein-Rosen bridges 
implies that ER=EPR could in-principle be tested by observers who themselves  never cross the horizon. The point is not that \qt  \it cannot \rm be understood without wormholes, but that it \it can\rm \ be understood (geometrically) with wormholes\footnote{Maldacena, Stanford, and Yang expressed it this way: ``What is interesting is not so much that information can be transferred, since after all
we are explicitly coupling the two systems. What is interesting and surprising is how."}.
\bn

In this paper we will illustrate some protocols for quantum teleportation. Assuming the existence of gravitational duals, the protocols must have bulk descriptions involving traversable wormholes.

\section{ Dynamics}

The approach of \cite{Gao:2016bin} is to stay within the highly controlled framework of AdS/CFT and to work with systems which are known to have gravitational duals. The arguments for traversability are tight and give a proof of concept.

However, teleportation through  wormholes  is not dependent  on conformal symmetry, or on AdS boundary conditions; teleportation is possible for entangled black holes, or more generally entangled horizons, in any background. The approach of  \cite{Susskind:2014yaa} was based only on the existence horizon  microstates and the fast-scrambling properties of horizons. This paper follows the latter strategy.

We will assume that the microstates of the horizon are described by k-local dynamics which is needed for fast scrambling. Then, after constructing protocols for teleportation we will
 compare them with  \cite{Gao:2016bin} and \cite{Maldacena:2017axo}. 

\section*{\small Qubit Model}

The systems we will consider, including black holes, are the
\it   fast scramblers \rm \cite{Sekino:2008he}. Ordinary Schwarzschild black holes are fast scramblers and so are AdS black holes whose Schwarzschild radii are the same as the AdS length scale. Nothing new would be added by studying larger black holes in AdS.

Systems exhibit fast scrambling if they are governed by k-local, but not spatially local, dynamics. For simplicity we can choose $k=2.$
 In appropriate units the scrambling time for such Hamiltonians is $t_* = \log S $ where $S$ is the entropy of the system.
  The minimal number of qubits needed to describe or simulate  a system of  a given entropy   is the entropy itself. 
 A black hole of entropy $N$  may be approximately  described as $N$ qubits at infinite temperature. We write the Hamiltonian in the form,
\be 
H = \sum_{ij} h_{ij}
\label{H}
\ee 
 where $h_{ij}$ depends only on the Pauli operators of the $i^{th}$ and  $j^{th}$ qubits. 
 
\section*{\small Frozen Qubits}

This setup with a fixed number of qubits allows us to study a black hole in equilibrium, but typically, if we perturb the black hole its entropy will increase. In gauge-gravity duality the infinite collection of UV degrees of freedom serve as a reservoir of frozen or unexcited qubits which can be excited when energy is added the black hole. In the qubit model we can deal with this by
by assuming a collection of frozen  qubits.  Perturbations that increase the entropy of the black hole would be modeled by ``defrosting" qubits; that is, by turning on couplings  to the other interacting qubits.
 For example throwing a thermal photon into a black hole will increase the black hole's entropy by one unit, requiring one qubit to be defrosted. Frozen qubits will be especially important when teleporting relatively large systems as in appendix \ref{App: Large}.

 In the following section we will need to study systems with $N-1, \ N, \rm \ and \ \it N+1$ qubits. We may start with an $(N+1)$-qubit Hamiltonian of the form \ref{H}. To decrease the number of qubits by one, we select a qubit and simply delete all terms in \ref{H} containing that qubit. The inverse process takes place if we add a qubit to the system.
 
\section*{\small Precursors} 
 
\begin{figure}[H]
\begin{center}
\includegraphics[scale=.4]{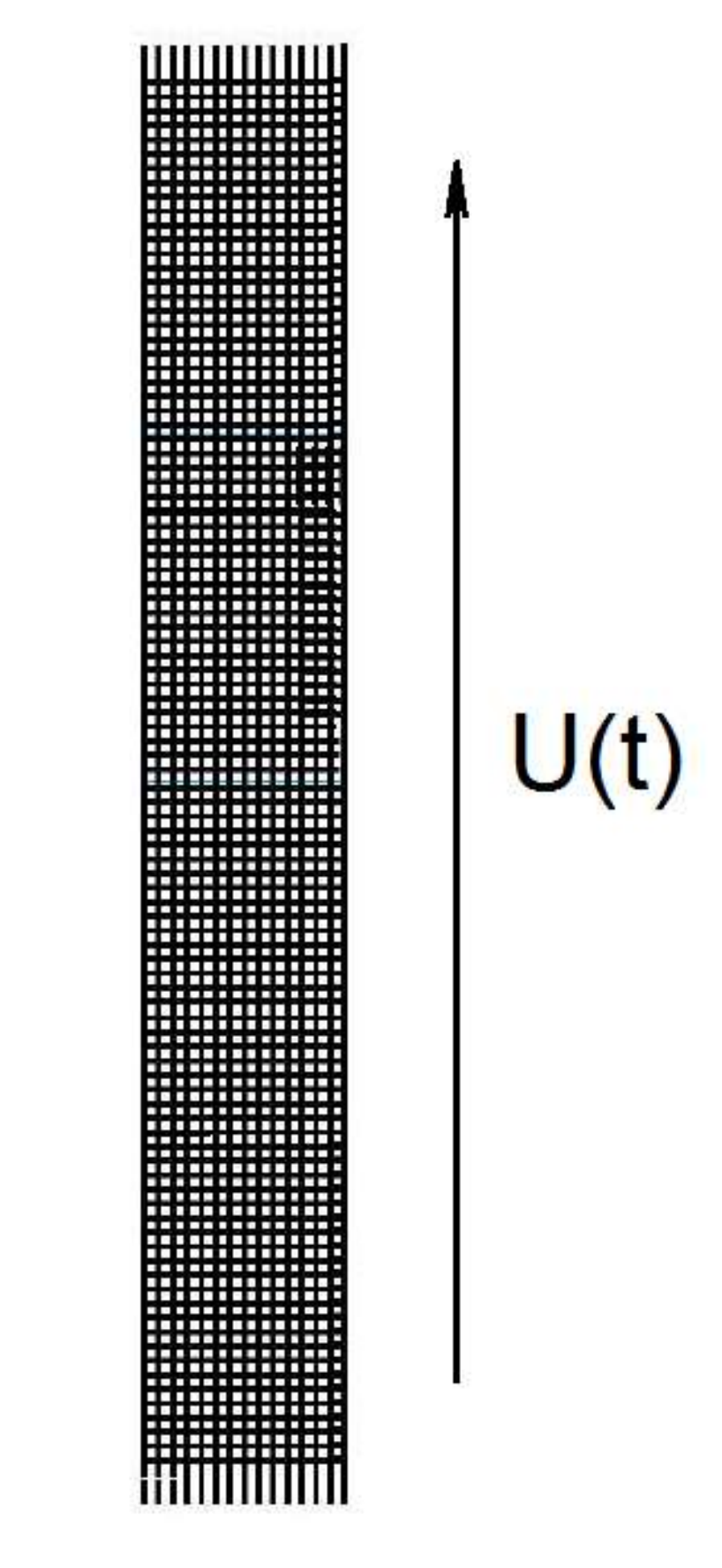}
\caption{Schematic illustration of a quantum circuit for the time-evolution operator $U(t).$}
\label{grid-minus1}
\end{center}
\end{figure}

The time evolution operator $e^{-iHt}$ of an $N$ qubit system may be thought of as a quantum circuit of width $N$ and depth equal to the amount of evolved time. We can schematically draw it  as in figure \ref{grid-minus1}  with the vertical lines representing  the $N$  qubits and the horizontal lines representing gates or the actions of the Hamiltonian. There is no significance to the regular lattice structure shown in the figure other than it is easy to draw.

\bn
Operators $A$ will be used to add or subtract a qubit from a circuit, thereby changing $N$. Pictorially, when $A$ acts a new qubit is added or removed from the circuit.

\begin{figure}[H]
\begin{center}
\includegraphics[scale=.4]{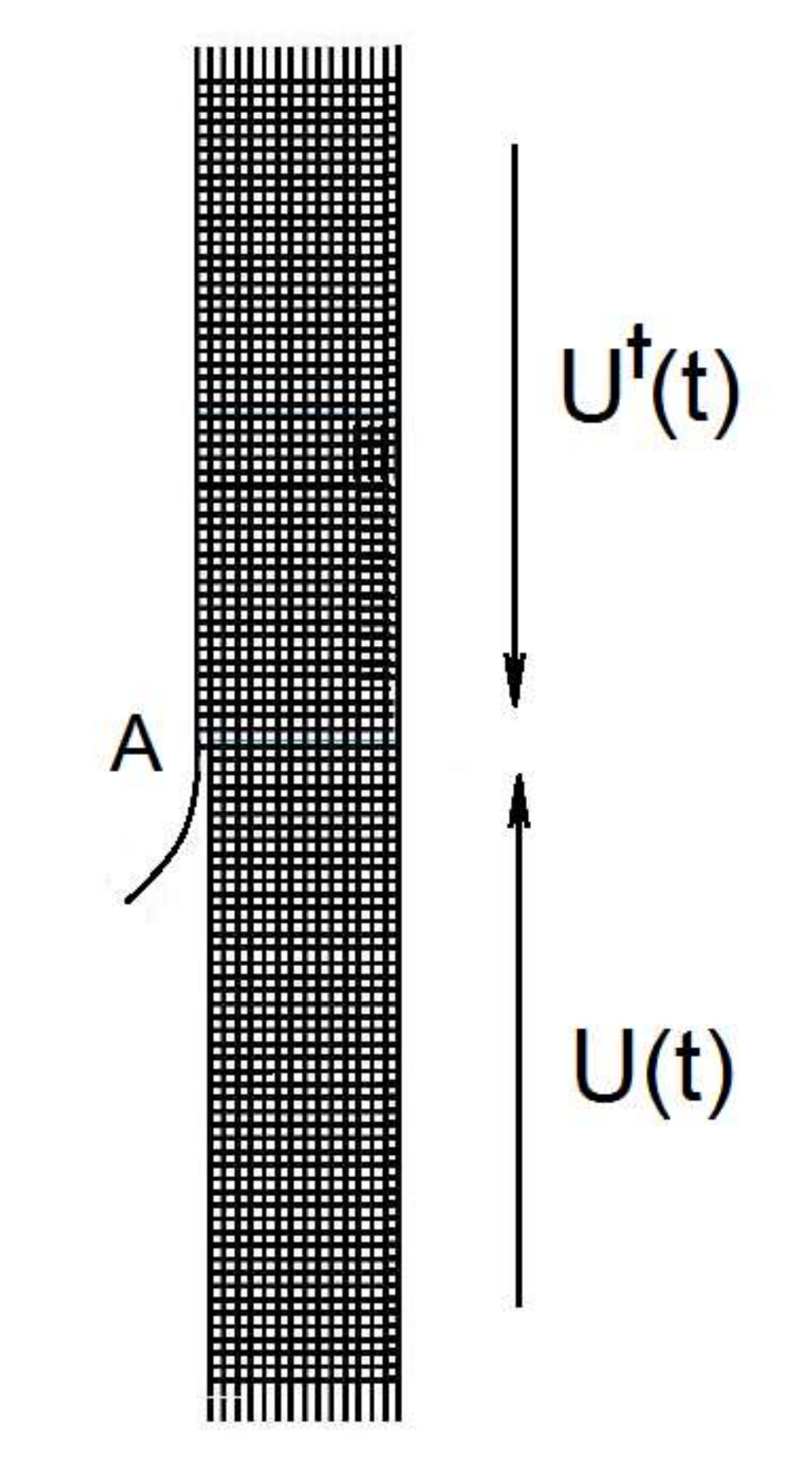}
\caption{Time evolution operator for a precursor $U^{\dag}(t) \  A \ U(t)$ Note that the operator $U(t)$ acts on $N$ qubits and $U^{\dag}(t)$ acts on $N+1$ qubits.  The arrows on the right indicate the direction of time-flow \cite{Heemskerk:2012mn}}.
\label{grid-0}
\end{center}
\end{figure}

An important class of operators are called precursors \cite{Heemskerk:2012mn}. Formally they look like Heisenberg operators but they are operators in the Schrodinger picture which are typically non-local and have a complexity which grows with the parameter $t$. The precursor associated with an operator $A,$ and a time $t,$ has the form,
 \be 
U^{\dag}(t) \ A \ U(t)
\label{precursor circuit}
\ee
 The circuit for such a precursor is shown in figure \ref{grid-0}

When calculating the complexity of a precursor there is a switchback effect which is due to cancellation between gates of $U$ and gates of $U^{\dag}.$ The switchback effect was described in  \cite{Stanford:2014jda}\cite{Susskind:2014jwa}\cite{Brown:2016wib}. We refer the reader to those papers and to Appendix \ref{size and complexity} for details.

\section*{\small Scrambling}

The unitary time evolution operator for a system of $N$ quits will be called $U_N(t).$ 
In appropriate units the scambling time for $k$-local quantum circuit is $t_* = \log N.$ The corresponding unitary time evolution operator $U_N(t_*)$ will be denoted by $V_N$ where $N$  labels the number of qubits in the circuit.
When there is no ambiguity we will drop the subscript $N$. 
\be
V \equiv e^{-iH t_*}
\ee

\begin{figure}[H]
\begin{center}
\includegraphics[scale=.3]{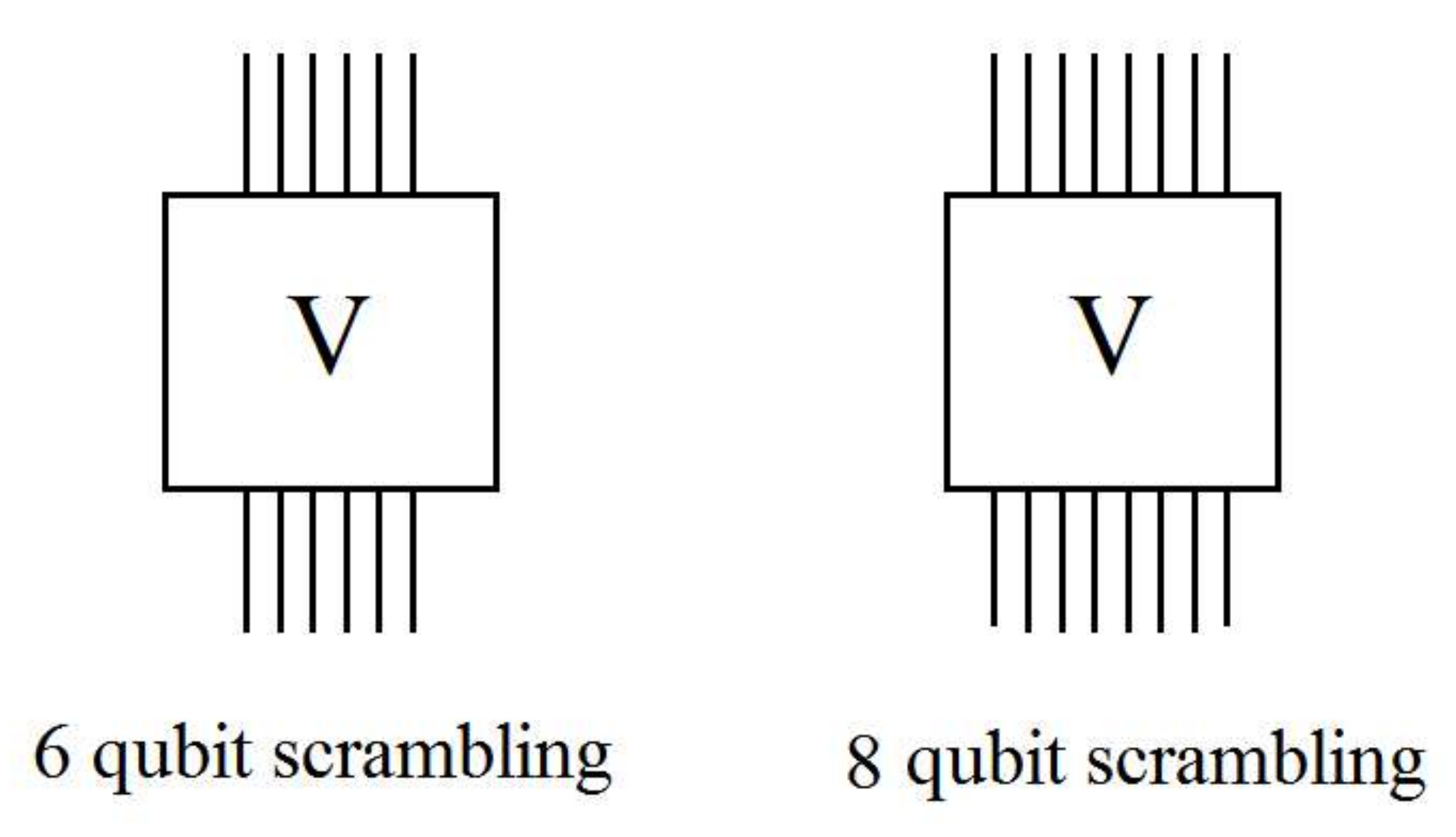}
\caption{Six and eight qubit scrambling operators.}
\label{n-qubits}
\end{center}
\end{figure}

 Graphically $V$  always has an equal number of in and out lines. If there are $K$ lines in, and $K$ lines out, then $V_K$ represents the evolution of an $K$ qubit system for a scrambling time. 
In all cases described in the subsequent circuit diagrams $K$ either equals $N$, $N+1$, or $N-1.$ The difference between scrambling times for these cases is negligible and can be ignored. In figure \ref{n-qubits} we illustrate the convention for $K=6$ and $K=8$.

We will be interested in the circuit complexity of the teleportation protocol and of the parts that comprise it. 
The circuit complexity of a scrambling operator $V$ is \cite{Brown:2016wib}
\be 
\CC(V) = N \log N.
\ee

We will also need to know the complexity of precursor operators $V^{\dag} W V$ where $W$ is a simple one or two qubit operator (in figure \ref{grid-0} the role of $W$ is played by the number-changing operator $A$ ). By the subadditivity of complexity, the complexity of the precursor is bounded by
\be 
\CC(V^{\dag} W V) \leq \CC(V^{\dag}) + \CC( W) + \CC( V) \approx 2N \log{N}.
\label{subadditivity}
\ee

However the switchback effect  \cite{Stanford:2014jda}\cite{Susskind:2014jwa}\cite{Brown:2016wib} \ (see also Appendix \ref{size and complexity}) \ 
implies that  $\CC(V^{\dag} W V)$ is much smaller than the bound in \ref{subadditivity} and is given by,
\be 
\CC(V^{\dag} W V)= N.
\label{switchback}
\ee

\section{Teleportation Protocol}\label{Sec: Protocol}

In this paper we will construct quantum-circuit protocols for teleportation through chaotic many-qubit \it mediators.\rm \  A discussion of complexity of decoding is in Appendix \ref{App: decodingcomplexity}. Assuming the existence of gravitational duals, we explain how teleportation may be related to the traversability phenomenon reported in \cite{Gao:2016bin} and elaborated in \cite{Maldacena:2017axo}.

\bn

Quantum teleportation involves three systems: 
two mediating entangled systems belonging to Alice and  Bob: and a third system  that Alice intends to teleport to Bob. We'll call the entangled system the \it mediator \rm and the system to be teleported, the \it teleportee. \rm Alice's portion of the mediator will be labeled $\bf A,$ Bob's portion by $\bf B,$ and the teleportee by $\bf T$.

The entanglement entropy of the mediator is $N.$ We  model it by  system of $2N$ qubits in a maximally entangled state. The teleportee will be modeled as a system of $n$ qubits in the state $|\Phi\ra$.
We will regard the teleportation as successful  if an $n$-qubit subsystem of Bob's share, materializes  in the pure state $|\Phi\ra$.

There are two facts about teleportation which we expect to be true for any teloportation protocol:

\begin{itemize}
\item  The process of teleportation always diminishes the amount of entanglement between Alice and Bob by at least $n$. Therefore the  teleportation can only be done if $n\leq N.$

\item{Quantum teleportation of an $n$ qubit system requires the transfer of at least $2n$ classical bits between Alice and Bob. The classical bit-transfer takes place through ordinary ``exterior" space-time.}

\end{itemize}

The special case in which $n=N$ (which we refer to as \it large-system-teleportation \rm) was studied in \cite{Susskind:2014yaa} and  reviewed in Appendix \ref{App: Large}.  Here we study the case $n<<N$  (\it small-system-teleportation \rm).   We will illustrate  small-system-teleportation for the case\footnote{In the context of black holes a one-qubit  teleportee may be taken to be a single quantum ( a graviton or a photon) of thermal energy. For a Scharzshild black hole this means  an  energy 
$$\delta E = \frac{1}{8\pi M G}.$$ The corresponding entropy added to Alice's black hole is $\delta S = 1$.  } $n=1$.

 The computational basis for an $N$-qubit system is defined as the simultaneous eigenvectors of the $Z$ Pauli operators $Z_i$ where $i$ runs from $1$ to $N$. A typical computational state of $N$ qubits will be labeled $|I\ra = |00101....10101\ra.$  For simplicity we will assume all measurements are in the computational basis. If we wish to make a measurement in some other basis we must first apply a unitary operator to bring the relevant operator to the computational basis and then perform the measurement.

Let us suppose that 
 the initial state of the mediator is  maximally entangled in the infinite temperature Thermofield-Double (TFD) state, 
 \be
|TFD \ra = \sum_I |I\ra_A  \ |I\ra_B
\label{TFD}
\ee
where $|I\ra_{A,B}$ indicate a complete set of states in the computational bases of Alice's and Bob's systems. The state \ref{TFD} is equivalent to a produt of $N$ Bell pairs,
\be 
|TFD\ra = \{ |00\ra + |11\ra \}^{\otimes N} 
\label{N-Bell}
\ee

Alice is also  in possession of the teleportee, in this case a $1$ qubit system in the pure state 
\be 
|\Phi\ra_T = \sum_{k}  \Phi(k) \  |k\ra_T.
\ee
Here $k$ is a complete set of computational states in the Hilbert space of the teleportee. In the case $n=1,$ $|k\ra$ takes on  two values 
$|0\ra$ and $|1\ra$.

The initial state is,
\be 
|initial\ra = \sum_{Ik} \  \Phi(k) \  |k\ra_T  \ |I\ra_A  \ |I\ra_B.
\label{initial}
\ee

The goal is to teleport the state $|\Phi\ra$ from Alice to Bob. This means that a subsystem of Bob's qubits (a single qubit in this case) is made to appear in the pure state 
\be 
|\Phi\ra_B = \sum_{j}  \Phi(j) \  |j\ra_B.
\ee

\section*{\small A Trivial Protocol}

\begin{figure}[H]
\begin{center}
\includegraphics[scale=.4]{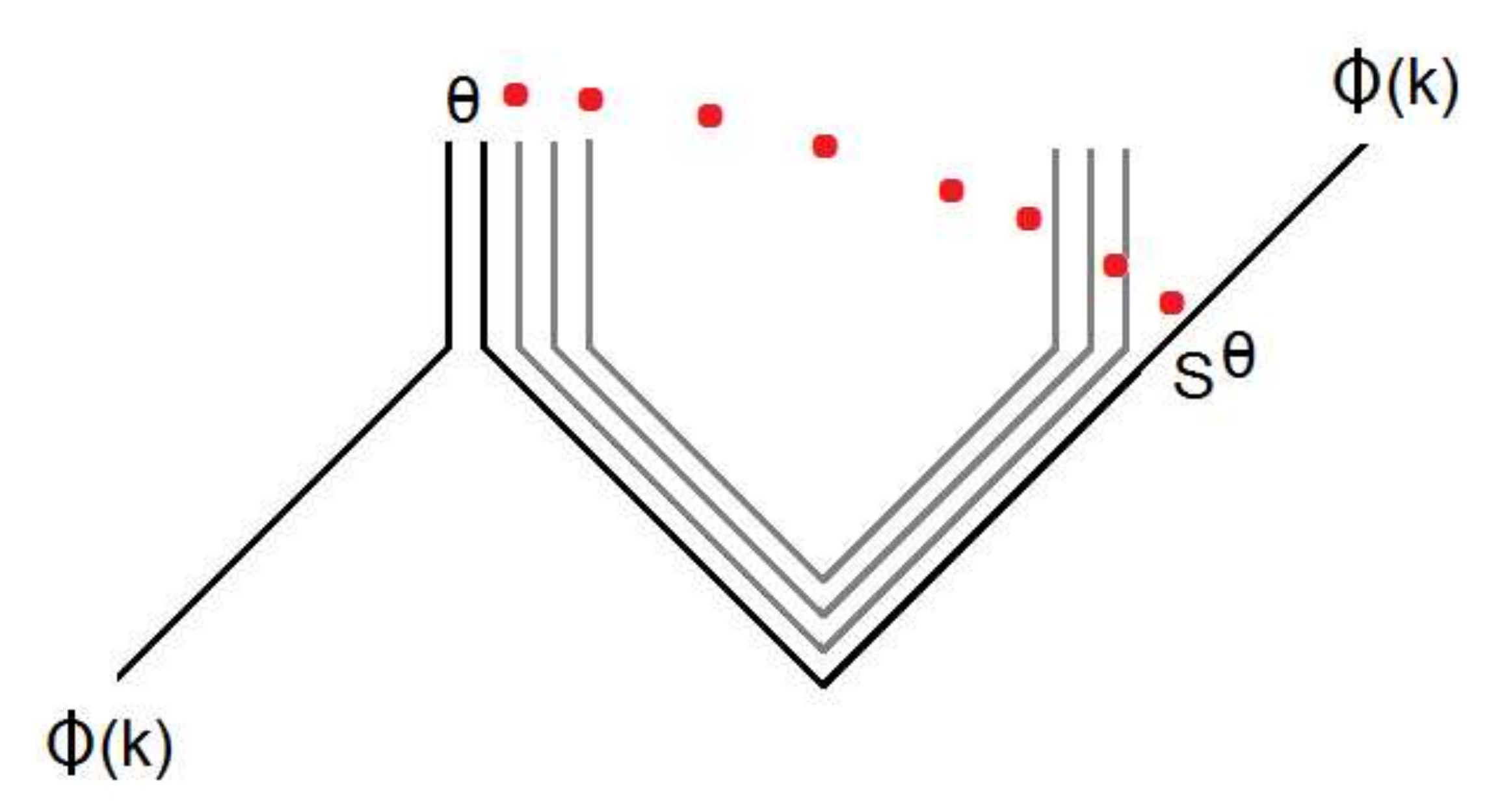}
\caption{A trivial way to teleport through the TFD state: The teleportee qubit is combined with one of the entangled qubits and a measurement is carried out in the Bell basis $|\theta\ra.$ The result is sent to Bob by classical communication of two bits. Bob applies the single qubit operator  $S^{\theta}$ and recovers the teleportee. }
\label{trivial}
\end{center}
\end{figure}

There is a trivial way to teleport a single qubit through a mediator in the state \ref{N-Bell}. We can disregard $(N-1)$ of the Bell pairs and just use the remaining Bell pair to do ordinary teleportation of the message qubit. This is illustrated in figure \ref{trivial}. 

\section*{ \small A More Interesting Protocol}

What is lacking in this protocol is any idea that the message was absorbed into Alice's black hole before being teleported. In order to avoid this trivial kind of teleportation we will assume that before the teleportation protocol begins, the teleportee-qubit is absorbed and scrambled with the black hole degrees of freedom. In other words 
 the teleportation protocol is delayed by a scrambling time $t_* = \log N$, during which time the dynamics of the black hole brings the teleportee and Alice's black hole into thermal equilibrium. 
 This may be represented by acting with a scrambling operator  on Alice's side immediately after the teleportee has been added to Alice's share of the mediator.

Starting with the initial state \ref{initial},  we combine Alice's share of the mediator with the teleportee to form a system $(\bf AT)$, and then apply the $(N+1)$-qubit scrambling operator $V_{N+1}$,
 \be 
V|initial\ra = \sum_{k,I} \Phi(k) V|k I\ra_{AT} |I\ra_B.
\ee

The next step in the protocol is to pick any two qubits from the $(\bf AT) $ system and label them $\theta.$ The notation $\theta$ will stand for a number of things; first of all it represents the two chosen qubits. In the form $|\theta\ra$ it represents the basis states for the two-qubit system in the computational basis. And finally it can stand for a bit string of length two representing the outcome of an experiment on the two qubits. The remaining $(N-1)$  qubits are labeled in a sumilar manner by  $\alpha.$ The $(\bf AT)$ system is the union of $\theta$ and $\alpha.$

The $(N+1)$-qubit scrambler $V$ acts giving,
\be 
 \sum_{k,I, \theta, \alpha} \Phi(k) \ |\theta, \alpha \ra \la \theta, \alpha | V  |k I\ra_{AT} |I\ra_B |
 \label{scrambled}
\ee
where $|\theta \ra$ and $|\alpha \ra$ are two-qubit and $(N-1)$-qubit computational states. 

We define the symbol $V^{\theta, \alpha}_{kI}$ by,
\be 
V^{\theta, \alpha}_{kI} \equiv \la \theta, \alpha | V  |k I\ra
\ee
 so that \ref{scrambled} becomes,

\be 
 \sum_{k,I, \theta, \alpha} \Phi(k) \ V^{\theta, \alpha}_{kI} \ |I\ra_B \ |\theta, \alpha\ra
 \label{scrambledegg}
\ee

Next, Alice measures the two qubit system $\theta$ and gets a particular outcome. She then sends the result, in the form of a classical two-bit string, $\theta$, to Bob.
Once Bob receives the message he then acts on his system with a unitary operator $Z^{\theta}$ that depends on the outcome $\theta$ of Alice's measurement. For each two-bit string $\theta$ on Alice's side, $Z^{\theta}$ is  unitary in the space of  Bob's $N$ qubit system. 

\begin{figure}[H]
\begin{center}
\includegraphics[scale=.4]{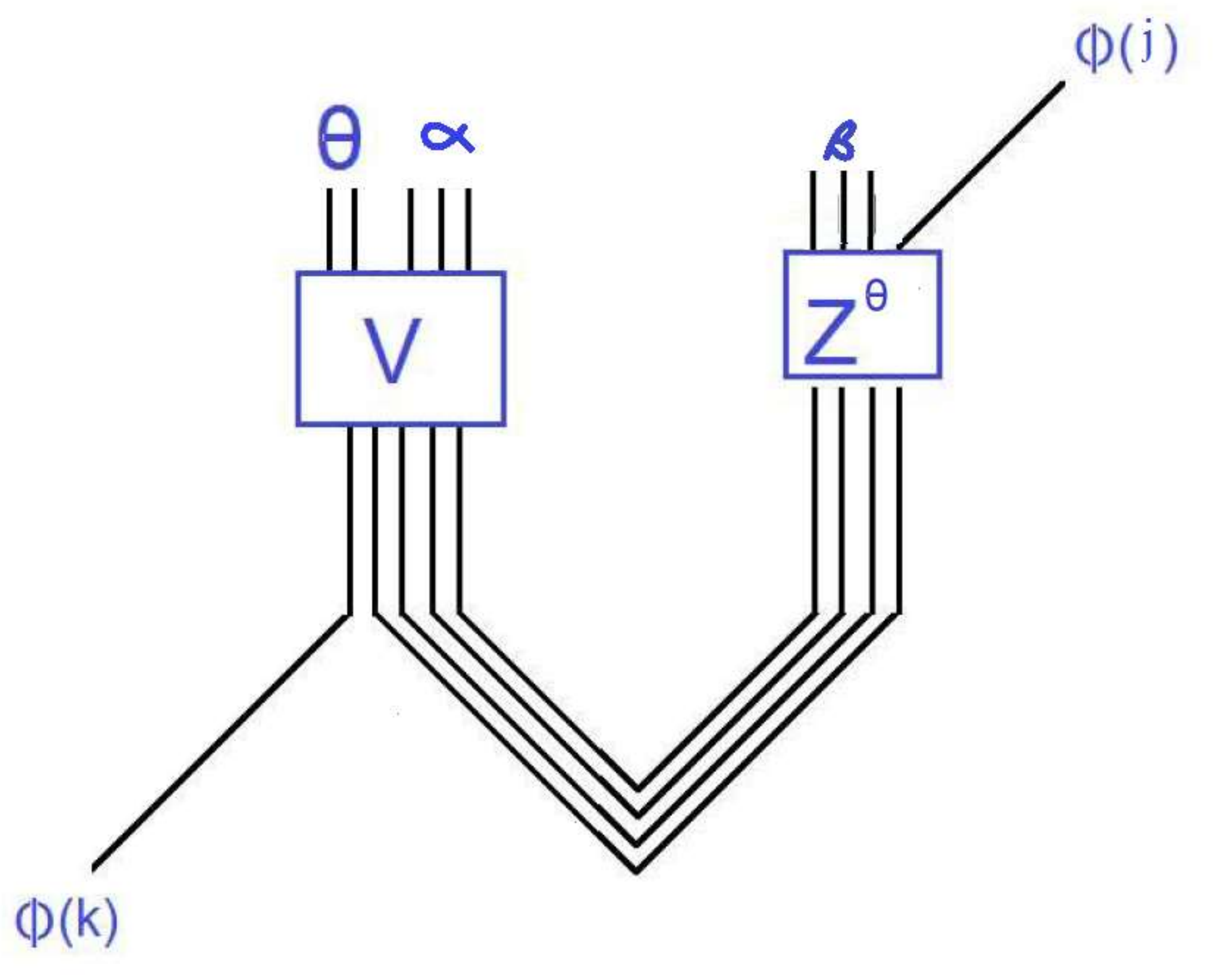}
\caption{A circuit for teleporting a single qubit through an entangled state of $N+N$ qubits.  At this stage the unitary operator  $Z$ on Bob's side is unspecified. }
\label{small-teleportation}
\end{center}
\end{figure}

The resulting state is 
\be 
 \sum_{k,I,  \alpha, j, \beta} \Phi(k) \ V^{\theta, \alpha}_{kI}  |\theta, \alpha\ra_{AT}   \  \la \beta, j| Z^{\theta}|I\ra_B \  |\beta, j\ra_B
 \label{scrambledegg}
\ee
 in which  the $N$ qubits on Bob's side have been partitioned into $(N-1)$ qubits labeled $\beta$ and a single qubit labeled $j.$ In figure \ref{small-teleportation} a circuit diagram is shown illustrating \ref{scrambledegg}.

Our goal is to show that $Z^{\theta}$ can be chosen (by Bob) so that after summation on $I$ the final state has the form,
\be  
\sum_{k, \alpha, \beta} \Phi(j) |\beta, j\ra_B \  W_{\alpha \beta} |\theta, \alpha\ra_A  
\label{W-eq}
\ee
where $W$ is a unitary matrix.  The meaning of this state is that it  consists of a qubit on Bob's side in the pure state $|\Phi\ra$ representing the succesfully teleported qubit;  a pair of qubits on Alice's side in the computational state $|\theta\ra$; and a maximally entangled system shared between Alice and Bob with entanglement entropy $(N-1).$ The maximally entangled system is in state $\sum_{\alpha \beta}  W_{\alpha \beta}|\alpha, \beta\ra.$

\begin{figure}[H]
\begin{center}
\includegraphics[scale=.4]{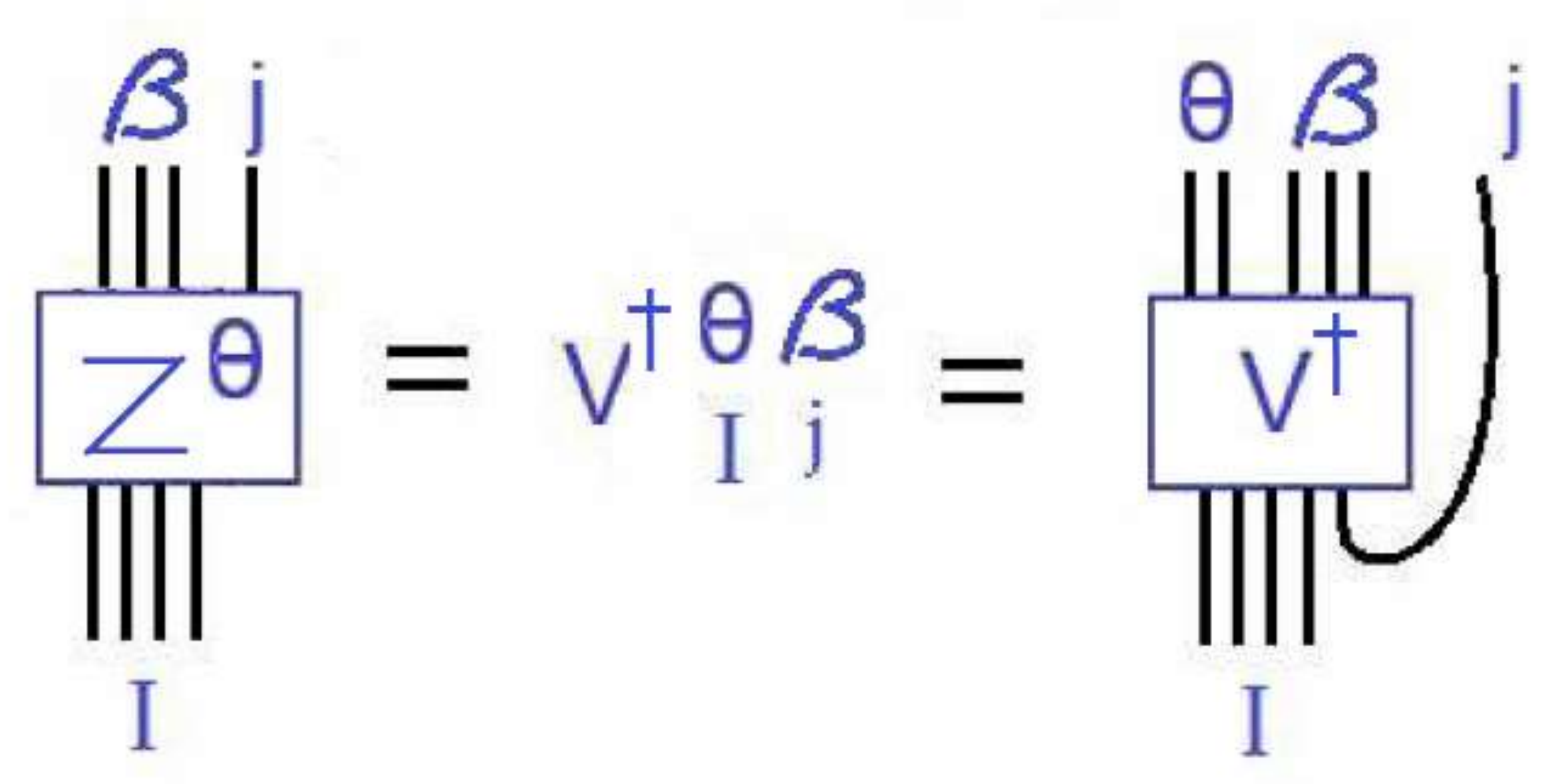}
\caption{Definition of the symbol  $V^{\theta,\beta}_{I,j}$. It may be thought of as a matrix connecting the $N$ qubit state $I$ to the $N$ qubit state $\alpha, j.$}
\label{crossing}
\end{center}
\end{figure}

To show this let us temporarily choose,
\be  
Z^{\theta}_{I, \beta j} = {V^{\dag}}^{\theta,\beta}_{I,j}
\label{temporary}
\ee
as shown in fig \ref{crossing}.

The important property of the  scrambling operator  $V$ is that it has a high degree of randomness although it is far from Haar random. One can say that it is 2-design random 
\cite{Hayden:2007cs}\cite{Roberts:2016hpo}. This is sufficient to show that for
 fixed $\theta,$ the matrix    $V^{\theta,\beta}_{I,j}$ is very close to a unitary matrix connecting the $N$-qubit states  $I$ and $\beta, j.$ Equation \ref{W-eq} then follows, but with $W_{\alpha \beta} = \delta _{\alpha \beta}.$ 
In other words the $\alpha,\beta$ system is left in the TFD state of $(N-1)$ qubit pairs.  

To get to a general $W$ we modify \ref{temporary} to,
\be  
Z^{\theta}_{I, \beta j} =\sum_{\gamma} {V^{\dag}}^{\theta,\gamma}_{I,j} \  W_{\gamma \beta}
\label{still-temporary}
\ee
Here ${V^{\dag}}^{\theta,\gamma}_{I,j}$ is the $(N+1)$-qubit inverse time evolution for time $t_*$ and $W$ is a matrix connecting $(N-1)$-qubit states $\gamma$ and $\beta.$

\section*{\small Minimizing the Complexity of Bob's Operation}

The choice of the matrix $W_{\alpha \beta}$ is arbitrary but we can use that freedom to make Bob's task as easy as possible. The difficulty of Bob's task can be quantified by the complexity of the operator $Z^{\theta}.$  The complexity of $V$ or $V^{\dag}$ is of order $N \log N.$ Generally the complexity of $Z$ in \ref{still-temporary} is bounded by,
\be 
\CC(Z) \leq   \CC(V^{\dag}) + \CC(W). 
\ee
which is at least $N\log{N}.$
But fortunately for Bob the switchback effect 
\cite{Stanford:2014jda}\cite{Susskind:2014jwa}\cite{Brown:2016wib} allows us to do much better than this bound. To see this choose, 
\be 
W_{\alpha \beta} = V^{\dag}_{\alpha \beta}
\label{W}
\ee
where $V^{\dag}_{\alpha \beta}$ is the inverse scrambling operator for $(N-1)$ qubits. 

\begin{figure}[H]
\begin{center}
\includegraphics[scale=.4]{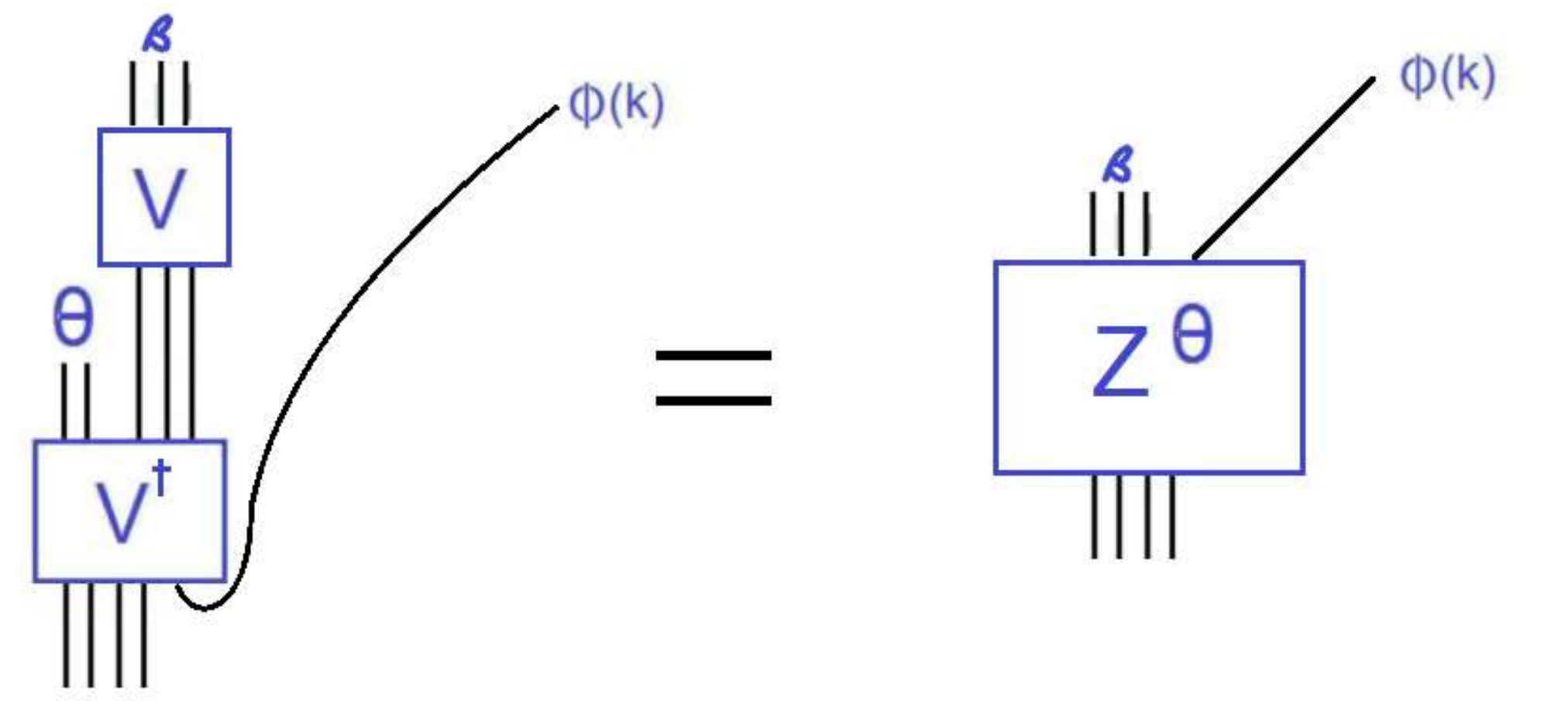}
\caption{Our choice of $Z^{\theta}.$ The complexity of $Z$ is only of order $N$ due to the switchback effect.}
\label{zee}
\end{center}
\end{figure}

With this choice  $Z$ takes the form,
\be  
Z^{\theta}_{I, \beta j} = \sum_{\gamma}V^{\theta,\gamma}_{I,j} \ V^{\dag}_{\gamma \beta}
\label{not-temporary}
\ee
A circuit diagram for \ref{not-temporary} is shown in figure \ref{zee}.

The product of a $V$ and a $V^{\dag}$  in figure \ref{zee} is very similar to a precursor $U(t) W U^{\dag}(t)$ where the time $t$ is the scrambling time $t_*$ and the insertion $W$ is replaced by the insertion of the two-qubit state $|\theta\ra.$
As in the precursor case we expect most of the complexity of $V$ and $V^{\dag}$ to cancel 
\cite{Stanford:2014jda}\cite{Susskind:2014jwa}\cite{Brown:2016wib} leaving a complexity of order $N$. Thus the complexity of $Z^{\theta}$ satisfies
\be 
\CC(Z^{\theta}) \approx N.
\ee
This is probably  the smallest complexity possible for $Z$.

\begin{figure}[H]
\begin{center}
\includegraphics[scale=.45]{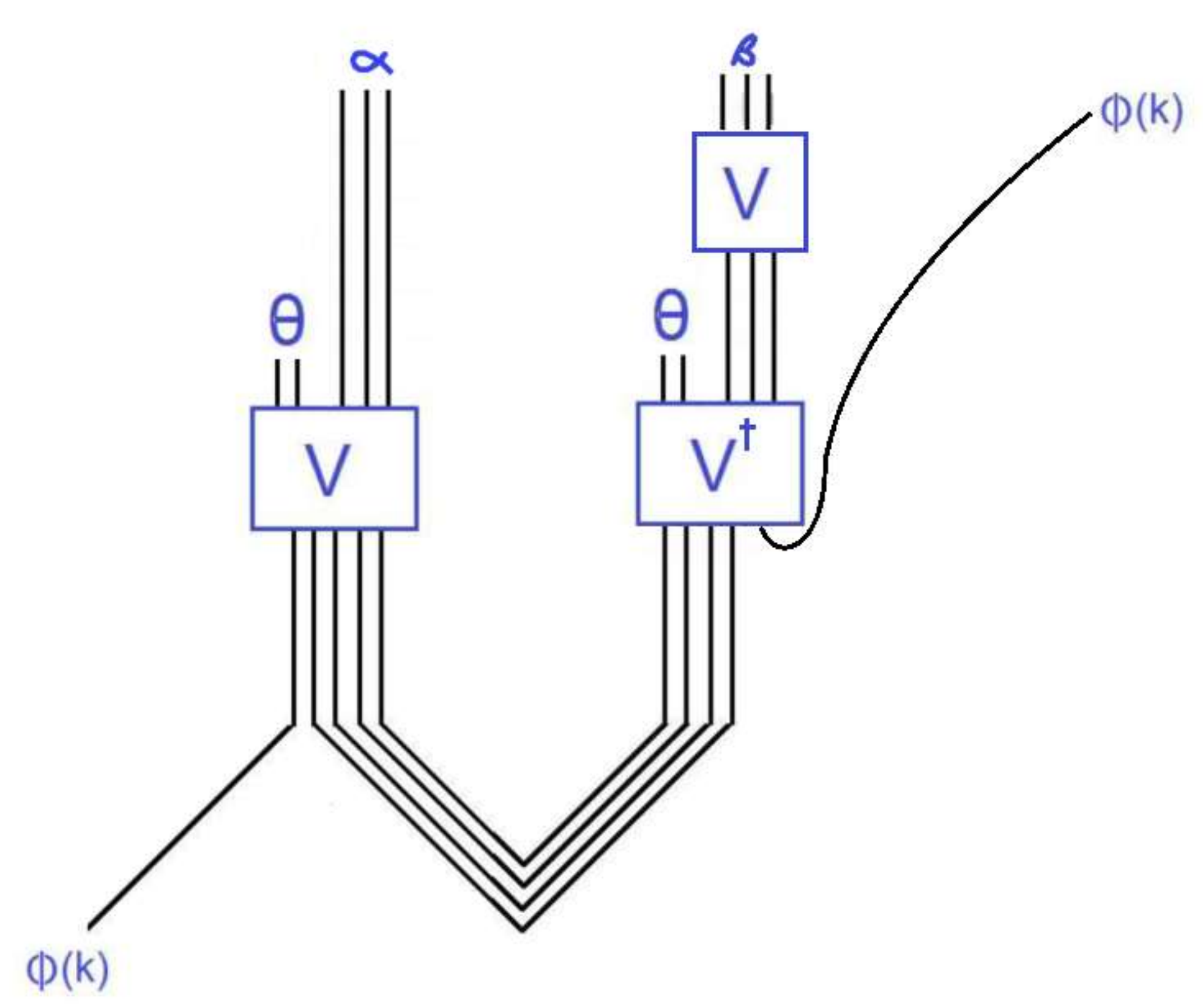}
\caption{ Teleportation circuit for acting on Bob's side at $t_R=0.$ }
\label{tR=0}
\end{center}
\end{figure}

In figure \ref{tR=0} the teleportation protocol of figure \ref{small-teleportation} is updated to account for the assumed form of $Z$,

\begin{figure}[H]
\begin{center}
\includegraphics[scale=.4]{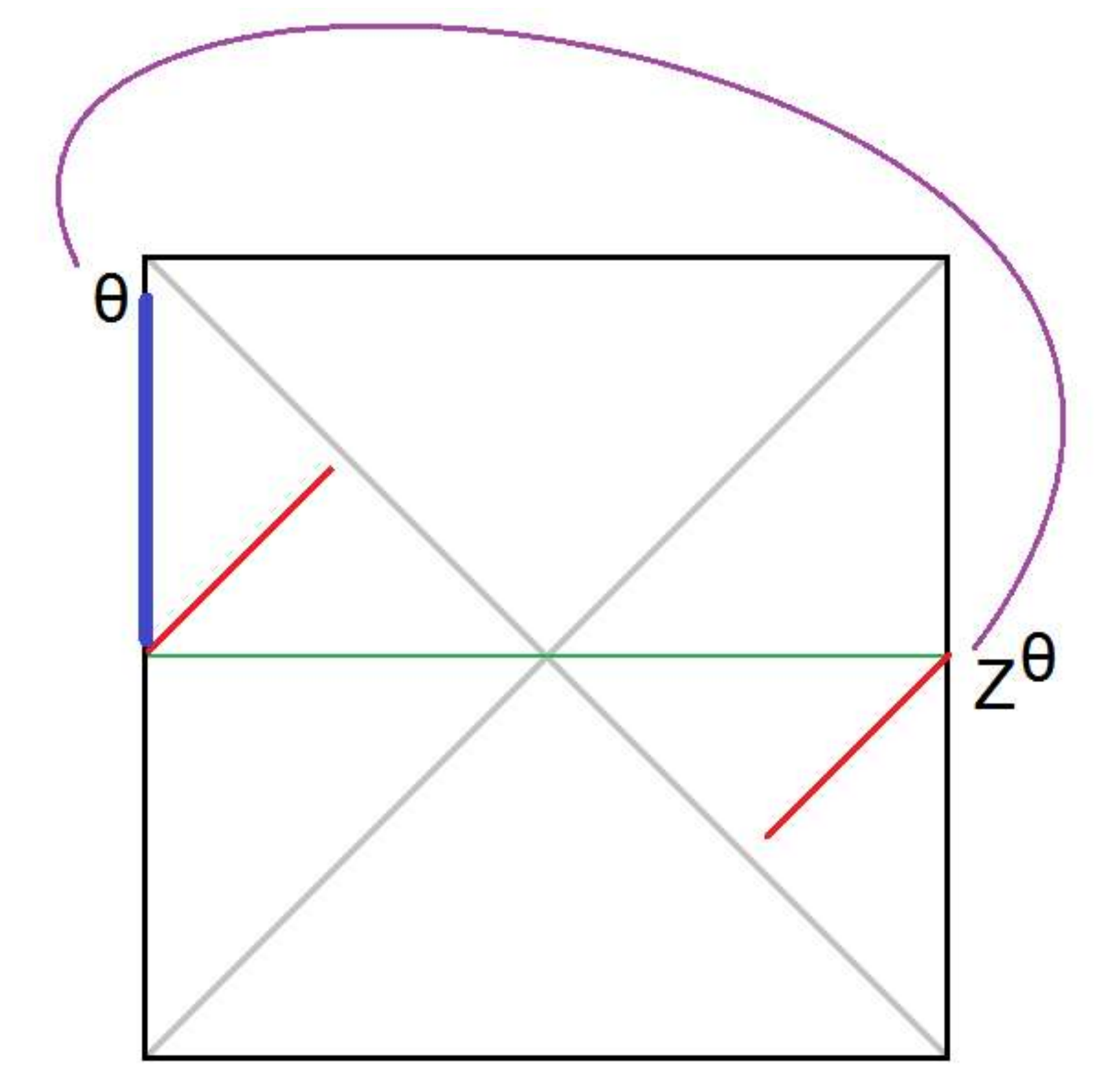}
\caption{Alice measures two qubits and sends the output $\theta$ to Bob in the form of a classical 2-bit message. Bob applies the unitary $Z^{\theta}$ at $t=0.$ to recover the message qubit on his side.}
\label{Penrose4}
\end{center}
\end{figure}

Going back to figure \ref{small-teleportation} the operator $V$ represents the forward time evolution of Alice's system from $t=0$ to $t=t_*.$ On Bob's side we have not accounted for a similar evolution. One interpretation is that the protocol requires the evolution on Bob's side to be temporarily frozen while Alice's side evolves to the scrambling time. We will think about it differently. Instead of Alice sending the bit-string $\theta$ at time $t_*$ and having Bob receive it at $t_*$ we will suppose that Bob receives it at time $t=0.$
In figure \ref{Penrose4} a Penrose diagram is used to illustrate the protocol.

\bn
Sending the message $\theta$ backward in time may seem unphysical but it is just a temporary trick that we will eventually not need. For the moment we will allow it.

\begin{figure}[H]
\begin{center}
\includegraphics[scale=.4]{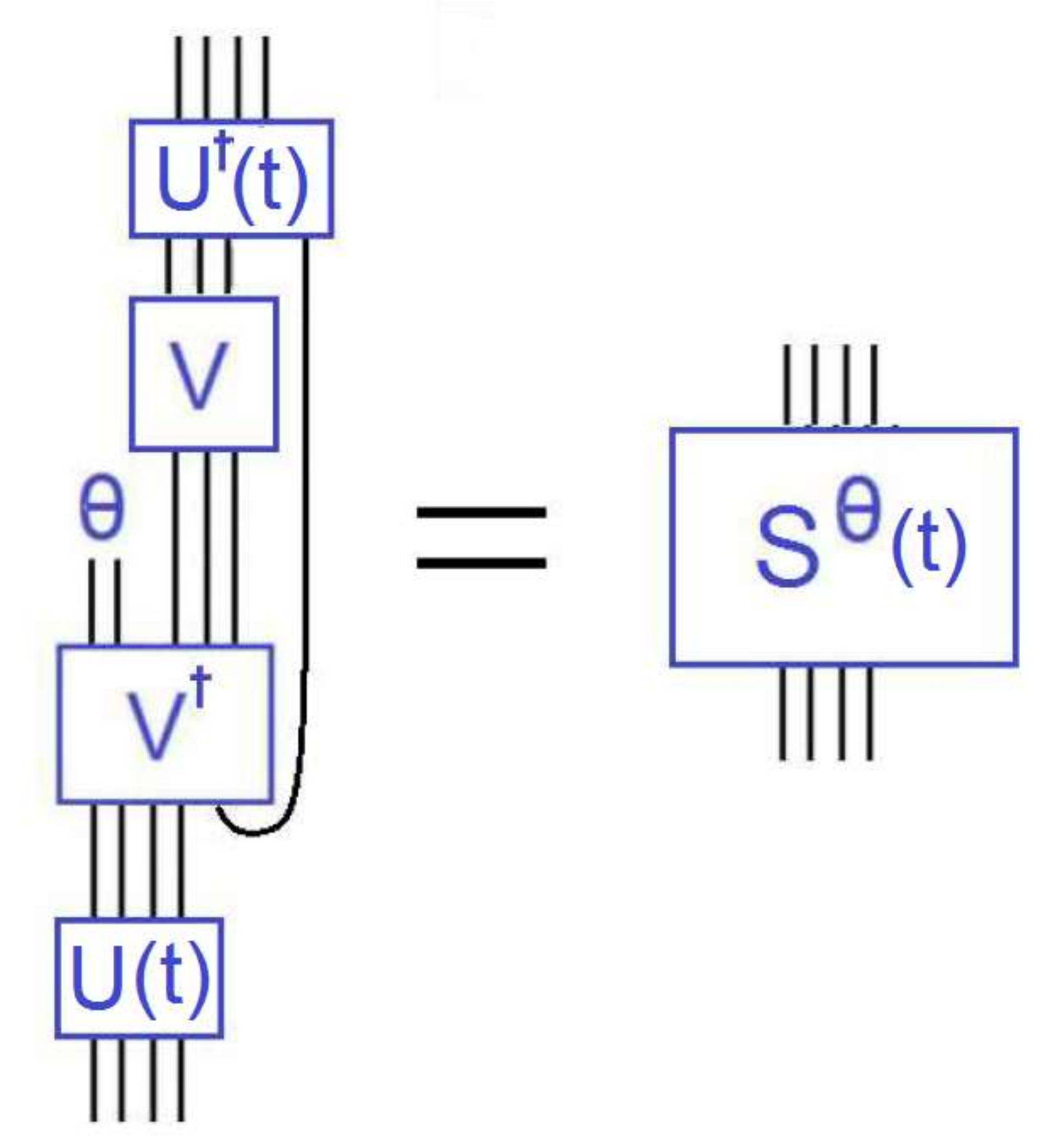}
\caption{Definition of the $N$ qubit unitary operator $S^{\theta}(t).$ One of the qubits on the right side loops from the bottom of $V^{\dag}$ to the bottom of $U^{\dag}(t).$ We'll call it the ``looping qubit".}
\label{subfinal}
\end{center}
\end{figure}

In fact we can allow Bob to act at any past time $-t$  by using the  precursor trick. 
Define 
\be 
S^{\theta}(t) = U^{\dag}(t) Z^{\theta} U(t).
\label{S(t)}
\ee
The effect of applying $S^{\theta}(t)$ at time $-t$ is identical to the effect of applying $Z^{\theta}$ at $t=0.$ 
 The circuit for $ S^{\theta}(t)$ is shown in figure \ref{subfinal}.

\section*{\small Complexity of $\bf \rm S^{\theta}(t)$}

\begin{figure}[H]
\begin{center}
\includegraphics[scale=.65]{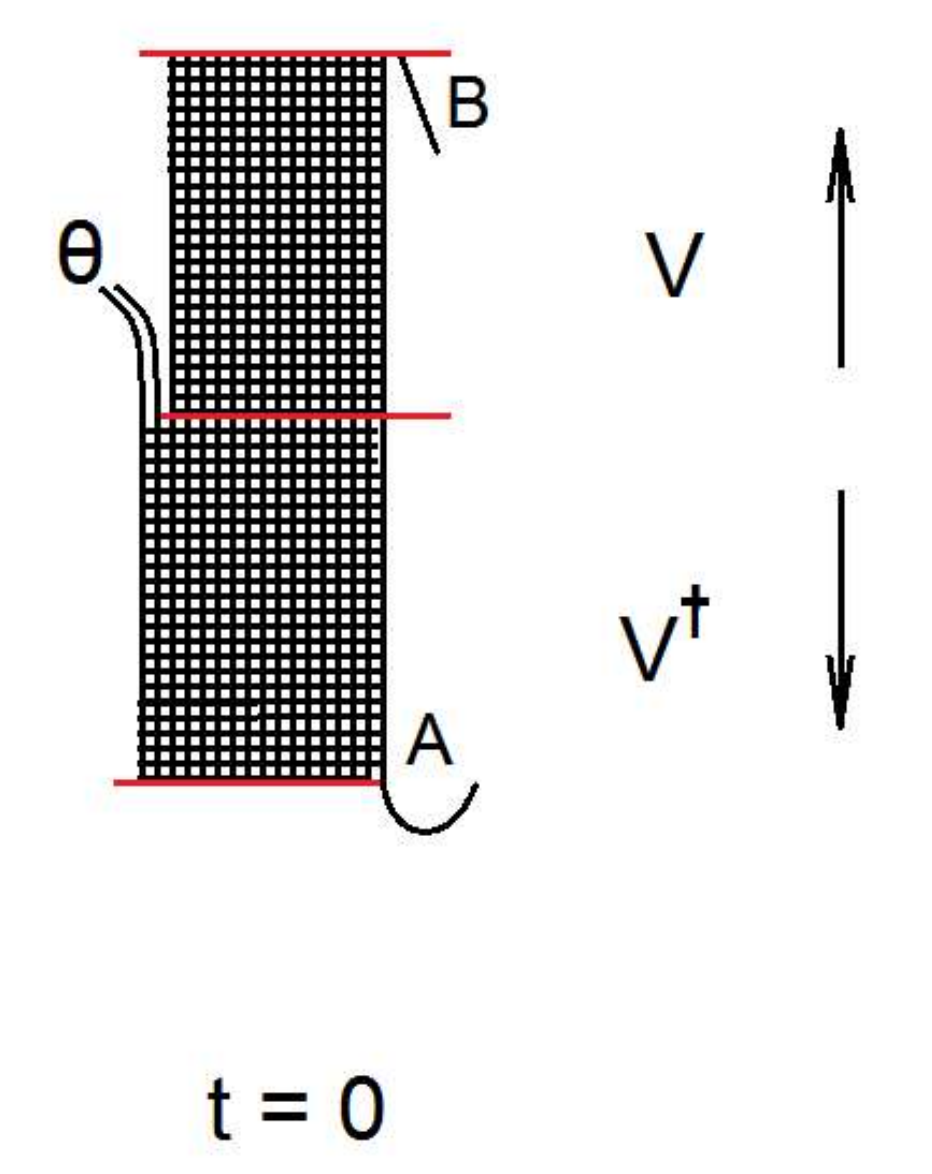}
\caption{Schematic circuit diagram for $S^{\theta}(0)$.}
\label{grid1}
\end{center}
\end{figure}

Later, in describing the gravitational dual of the teleportation protocol, we will need to know the complexity of the $S^{\theta}(t).$
First consider  $S^{\theta}(0)$ which is the same as $Z^{\theta}.$ We've already noted the similarity of this operator to a standard precursor evaluated at the scrambling time for which the complexity is $N$. To illustrate this point more graphically consider the  circuit in figure \ref{subfinal} in which we replace the boxes by many gates. This is shown in figure \ref{grid1} 
We have also included two insertions $A$ and $B$ associated with the looping qubit in figure \ref{subfinal}. We may think of the figure as describing an operator of the form
\be 
S^{\theta}(0) = B V \theta V^{\dag}A
\ee
The factors $A$ and $B$ each involve only a small number of qubits and carry complexity of order $1$. The rest of the circuit is a precursor and the switchback effect implies that it has complexity $N$.

\begin{figure}[H]
\begin{center}
\includegraphics[scale=.65]{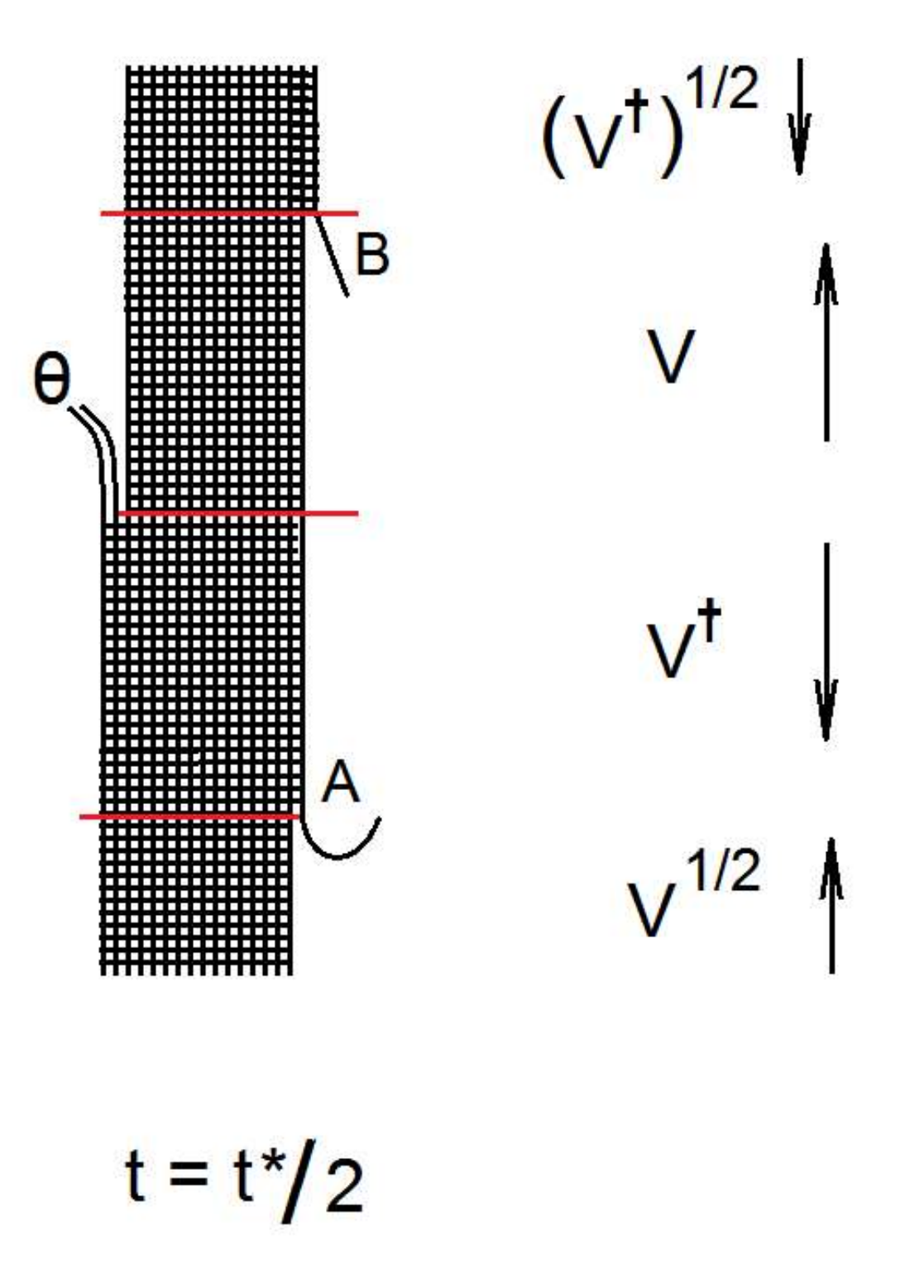}
\caption{Schematic circuit diagram for $S^{\theta}(t_*/2)$.}
\label{grid2}
\end{center}
\end{figure}

Next let us consider $S^{\theta}(t_*/2)$ for which the circuit is shown in figure \ref{grid2}
We may write the operator in figure \ref{grid2} in the form,
\be
(V^{\dag})^{1/2} \ B  \ V  \ \theta \  V^{\dag} \ A \  V^{1/2}  
\ee
which may be rewritten  as 
\be
S^{\theta}(t_*/2)=
\big\{(V^{\dag})^{1/2} \ B  \ V^{1/2} \big\} \ \big\{ V^{1/2}  \ \theta \ (V^{\dag})^{1/2} \big\} \big\{(V^{\dag})^{1/2} \ A \  V^{1/2} \big\}
\ee
This is a product of three precursors, each with a complexity of order $N^{1/2}.$ It follows that its complexity is bounded by order 
$N^{1/2}.$ In fact we don't find any evidence for cancellations which would make the complexity of $S^{\theta}(t_*/2)$ smaller than $N^{1/2}.$ Thus between $t=0$ and $t=t_*/2$ the complexity of $S^{\theta}$ decreases from $N$ to $\sqrt{N}.$

\begin{figure}[H]
\begin{center}
\includegraphics[scale=.65]{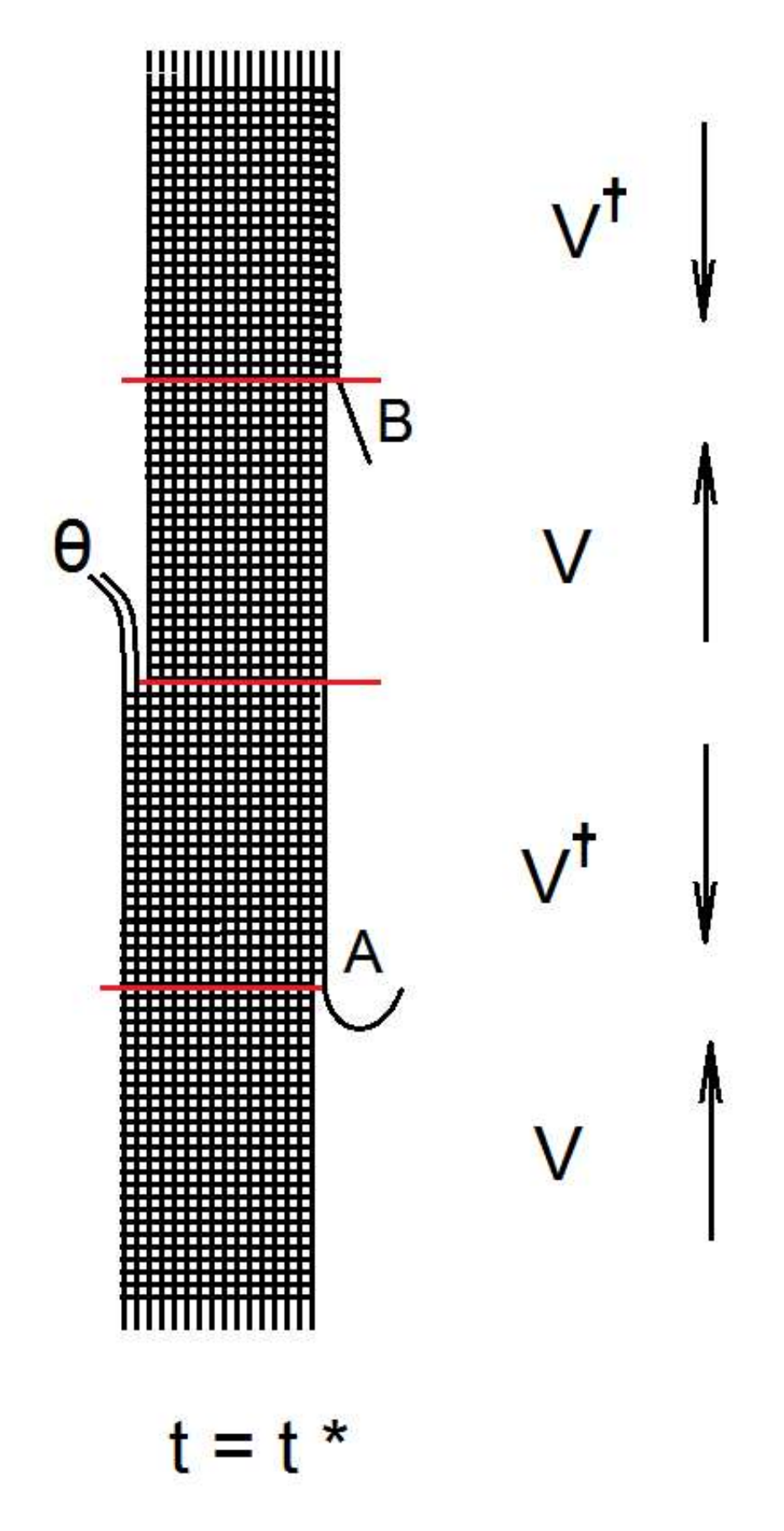}
\caption{Schematic circuit diagram for $S^{\theta}(t_*)$.}
\label{grid3}
\end{center}
\end{figure}

We might expect that as we increase $t$ toward $t_*$ the complexity of $S^{\theta}$ continues to decrease but that does not seem to be the case.  In figure \ref{grid3} a circuit for $S^{\theta}(t_*)$ is shown. 

Think of figure \ref{grid3} as the circuit for an operator 
\be 
S^{\theta}(t_*) = V^{\dag}B V \theta V^{\dag}AV.
\label{S=VdagBVthVdagAV}
\ee
Obviously,
\be 
S^{\theta}(t_*) =   (V^{\dag})^{1/2} S^{\theta}(t_*/2)   V^{1/2}   
\ee
By thinking of $S^{\theta}(t_*)$ as the precursor of  low size operator\footnote{ Size is being used in a technical sense here. See \cite{Size} and Appendix \ref{size and complexity}}  $S^{\theta}(t_*/2)$ (size of order $\sqrt{N}$) we can show, that in the absence of any further cancellation, that the complexity of $ S^{\theta}(t_*)$ is  $\sim N$. In fact it is the presence of the insertions $A$ and $B$ in \ref{S=VdagBVthVdagAV} that prevent us from collapsing $S^{\theta}(t_*)$ down to complexity of order $1$.

\begin{figure}[H]
\begin{center}
\includegraphics[scale=.4]{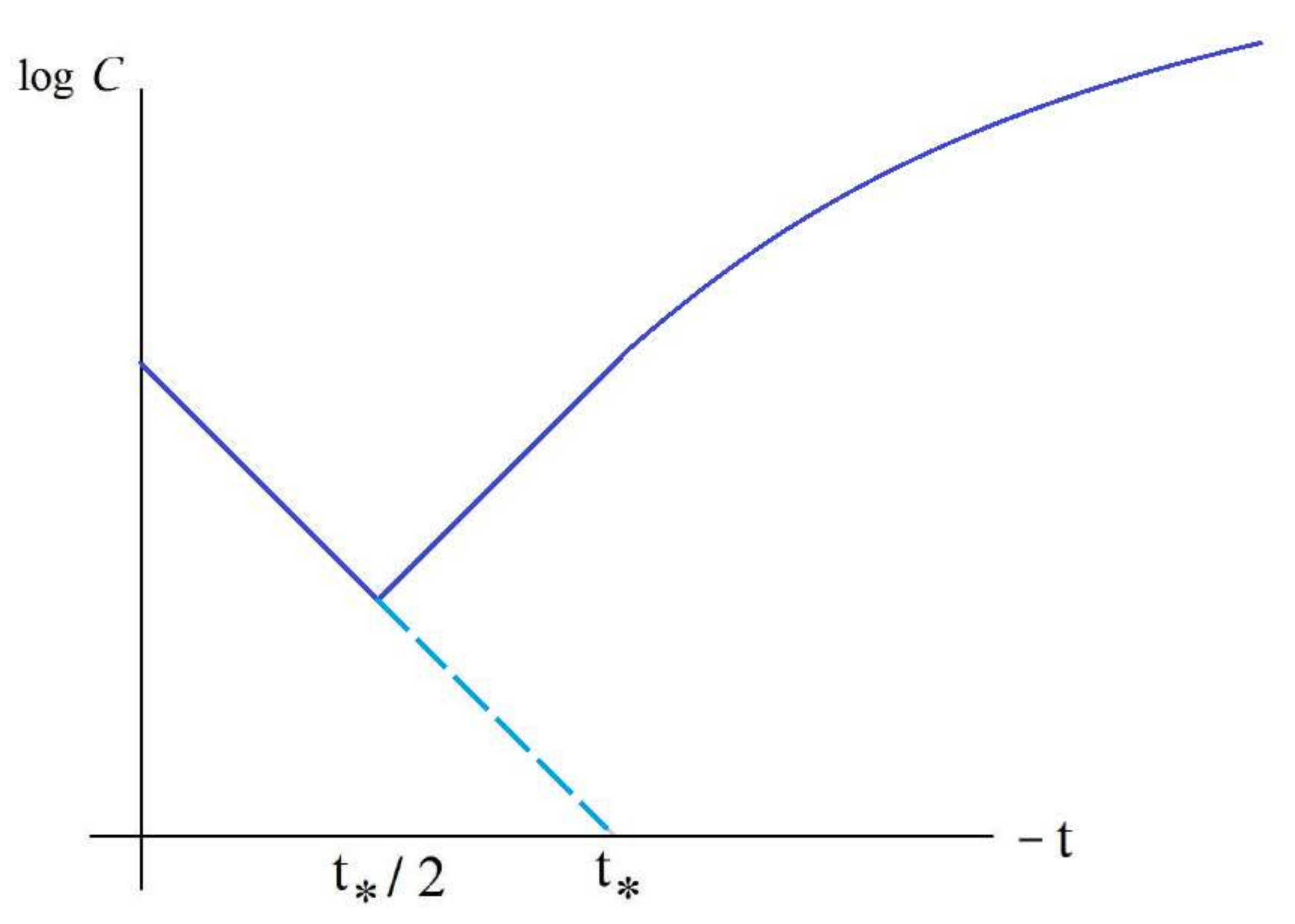}
\caption{The complexity of $S^{\theta}(t)$ reaches a minimum of $N^{1/2}$ at $t=- t_*/2.$ The light blue line shows the fictitious extrapolation of the complexity to $\CC \sim 1$ at $t=t_*.$}
\label{Complexity}
\end{center}
\end{figure}

By varying the time $t$ in \ref{S(t)} the complexity of the operator $S^{\theta}(t)$ can be varied. The complexity of $S^{\theta}(t)$ will be called $\CC(t).$  What we find is that as $t$ varies from $0$ to  half the scrambling time $t = t_*/2,$ the complexity varies as
\be  
\CC = N e^{-t} \ \ \ \ \ \  (  \ 0< t < t_*/2 \ )
\ee
It reaches a minimum of $\sqrt{N}$ at $t=t_*/2$ and then begins to grow as $t$ continues into the past. The behavior is summarized  figure \ref{Complexity}.

Later we will see that for some purposes we may pretend that the complexity continues to decrease to $\sim 1$ as $t$ tends to $t_*$ as indicated by the light blue broken line.


\section*{\small Unitary Operator Reformulation }

It is possible to repackage the teleportation protocol so that the process of measurement; the transfer of classical bits; and Bob's operation by $S^{\theta}$; are all replaced by a single unitary operator that couples the two sides, 
 
\be 
\CU(t) = \sum_{\theta}\Pi_{\theta}(t_*)\otimes S^{\theta}(t)   \ \ \ \ \ \ t<0.
\label{U}
\ee
$\Pi_{\theta}$  represents the projection operator onto the two qubit state $\theta,$  (tensored with the identity operator of the remaining system on Alice's side)  and $S^{\theta}(t)$ acts on Bob's side at time $-t.$   The sum involves four terms corresponding to $|\theta\ra = |00\ra,$ $|\theta\ra = |01\ra,$ $|\theta\ra = |10\ra,$ and $|\theta\ra = |11\ra.$

The result of carrying out the protocol is to leave the system at $t=0$ in the state,
\be 
\sum_{\theta}|\theta\ra_A \ \otimes \sum_{\alpha \beta} W_{\alpha \beta}|\alpha\ra_A | \beta \ra_B \ \otimes
\sum_j\Phi(j)|j \ra_B. 
\ee
In other words the two qubits $\theta$ on Alice's side are put into the pure state $|00\ra+|01\ra+|10\ra+|11\ra $, the remaining $(N-1)$
qubits $\alpha$ on Alice's side are maximally enangled with $(N-1)$
qubits $\beta$ on Bob's side, and what is most important, the one remaining qubit on Bob's side  has been teleported into the state $|\Phi\ra$.

Being unitary, the operator $\CU$ can be implemented by a direct coupling between the two sides as in 
\cite{Gao:2016bin}\cite{Maldacena:2017axo}.

\section{Gravitational Dual}

References  \cite{Gao:2016bin}\cite{Maldacena:2017axo} begin with a setup which is close enough to the usual AdS/CFT framework that it almost certainly has a gravitational dual, but it's relation to standard quantum teleportation is perhaps less obvious. By contrast we began with a quantum circuit description which is a generalization of standard quantum teleportation, but we cannot be sure that it has  a clean gravitational dual. We will suppose that the dual does exist and discuss its properties which we find to be similar but not identical to \cite{Gao:2016bin}\cite{Maldacena:2017axo}.

\begin{figure}[H]
\begin{center}
\includegraphics[scale=.4]{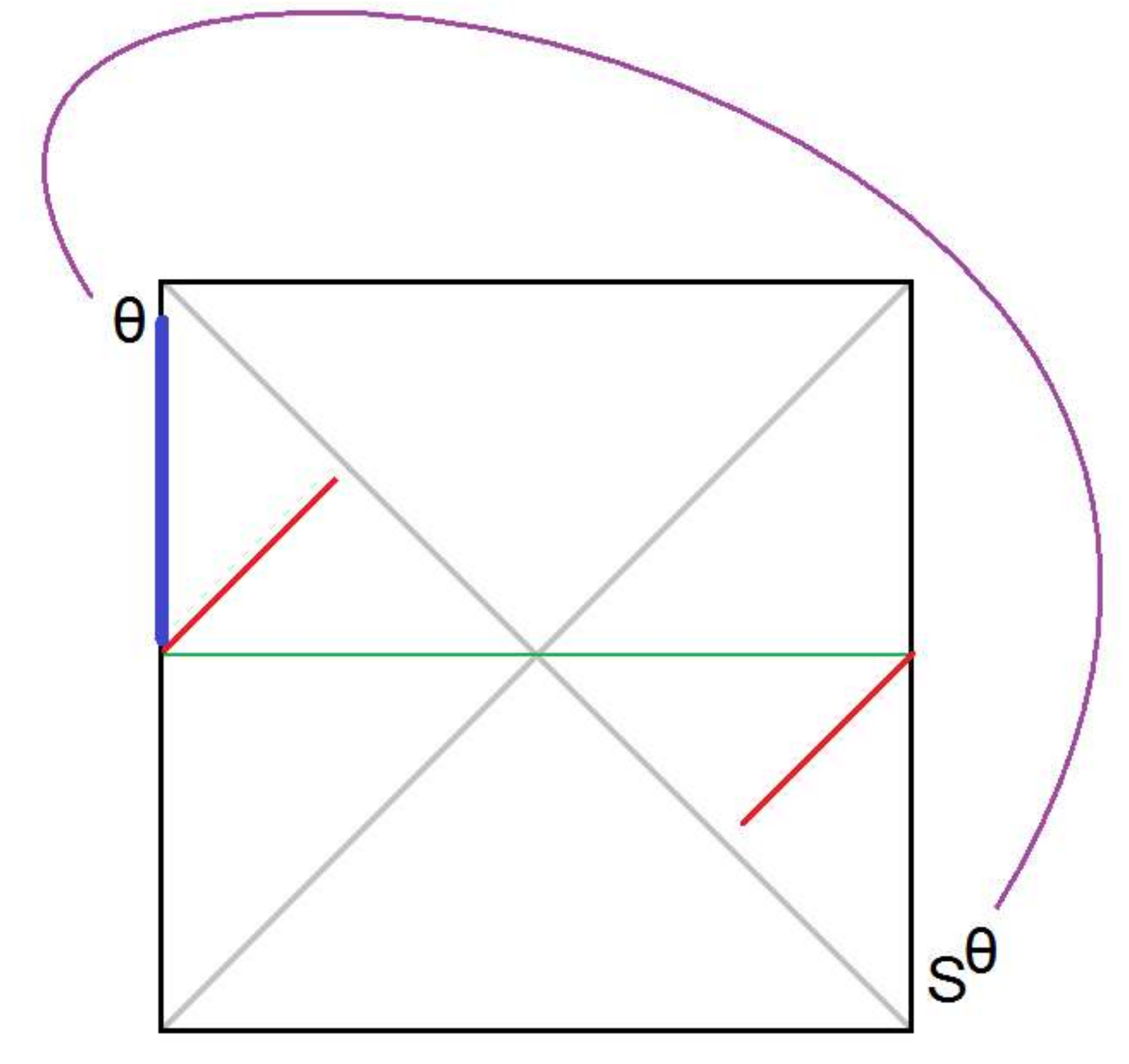}
\caption{ Bob can accomplish the same thing by applying $S^{\theta}$ at an earlier time $-t_*.$}
\label{Penrose5}
\end{center}
\end{figure}

Let's go back to figure \ref{Penrose4} which shows a Penrose diagram for our protocol in which Bob receives Alice's classical message at 
$t=0.$ A similar picture in figure \ref{Penrose5} shows the same protocol but with Bob receiving the message at a time $t<0.$ By applying $S^{\theta}(t)$ the same outcome is achieved.

\begin{figure}[H]
\begin{center}
\includegraphics[scale=.4]{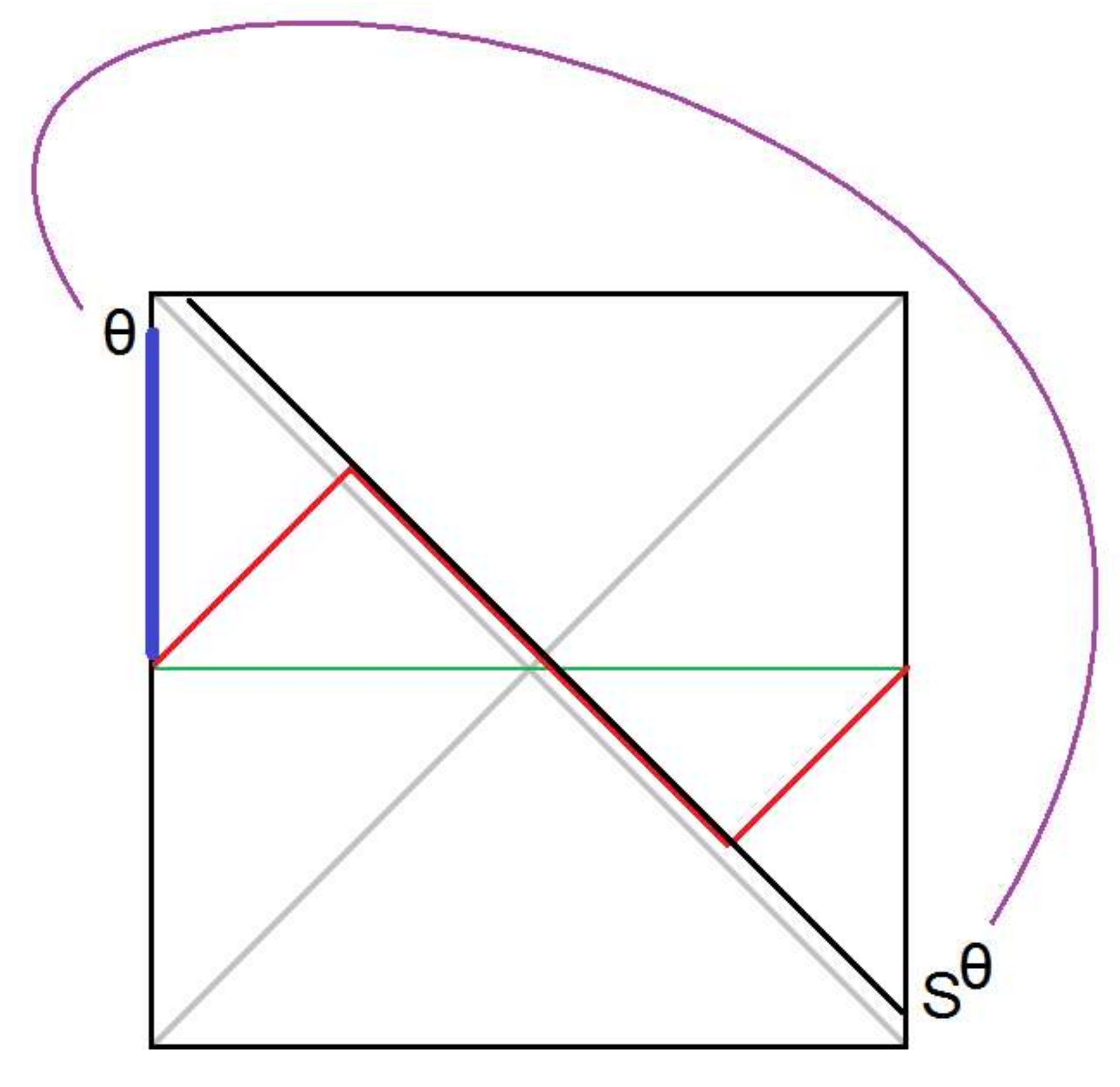}
\caption{Assuming $S$ is a simple operator of relatively low complexity,
it sends out a shock wave that interrupts the signal and shifts it back in time so that it comes out at $t=0.$}
\label{Penrose6}
\end{center}
\end{figure}

If $t$ is far in the past the effect of applying a perturbation will be to create a Stanford-Shenker shockwave traveling upward and to the left of the Penrose diagram as in figure \ref{Penrose6}.
Such a shockwave will have an effect on the trajectory of the teleportee shown in red in the earlier figure. If the shockwave is an ordinary positive energy shockwave it will cause the red trajectory to shift upward to the left so that it will fall into the singularity. 
However, our protocol was engineered to  ensure that the teleportee appears on Bob's side at $t=0$. There is only one way that can happen: as explained in \cite{Gao:2016bin}\cite{Maldacena:2017axo} the shockwave must have negative energy and shift the teleportee downward, to the right. This is shown in figure \ref{Penrose6}. 

The effect of the shock wave on the teleportee trajectory is parameterized by a single number that Shenker and Stanford call $\alpha$ \cite{Shenker:2013pqa}. It parameterizes the sign and magnitude of the shift and is only a function of the time $t$ that $S^{\theta}$ acts, and the energy carried by $S^{\theta}$. It satisfies,
\be
\alpha \propto E \ e^{|t|}
\ee
Depending on the sign of the energy the the shift may be positive or negative.

On the other hand the  complexity is linear in the magnitude of the energy, so the effect of the shock wave is proportional to 
\be
\alpha \propto \pm \CC \ e^{|t|},
\ee
the plus sign for positive energy shock perturbations and the minus sign for negative energy perturbations.

Now let us consider the case 
\be 
t = t_*/2, \ \ \CC = \sqrt{N}.
\ee
Using the fact that $t_* = \log{N}$ we see that this has exactly the same effect as a perturbation of unit complexity applied at time $t=-t_*.$
Thus, as far as the trajectory of the teleportee is concerned, we may pretend that $S^{\theta}(t_*)$ has complexity $1$ and apply it at $t=-t_*.$  To put it another way, we may pretend that 
in figure \ref{Complexity} the broken-line extrapolation is actually the correct behavior for $t_*/2 \ < t \ <  \ t_*.$

\begin{figure}[H]
\begin{center}
\includegraphics[scale=.4]{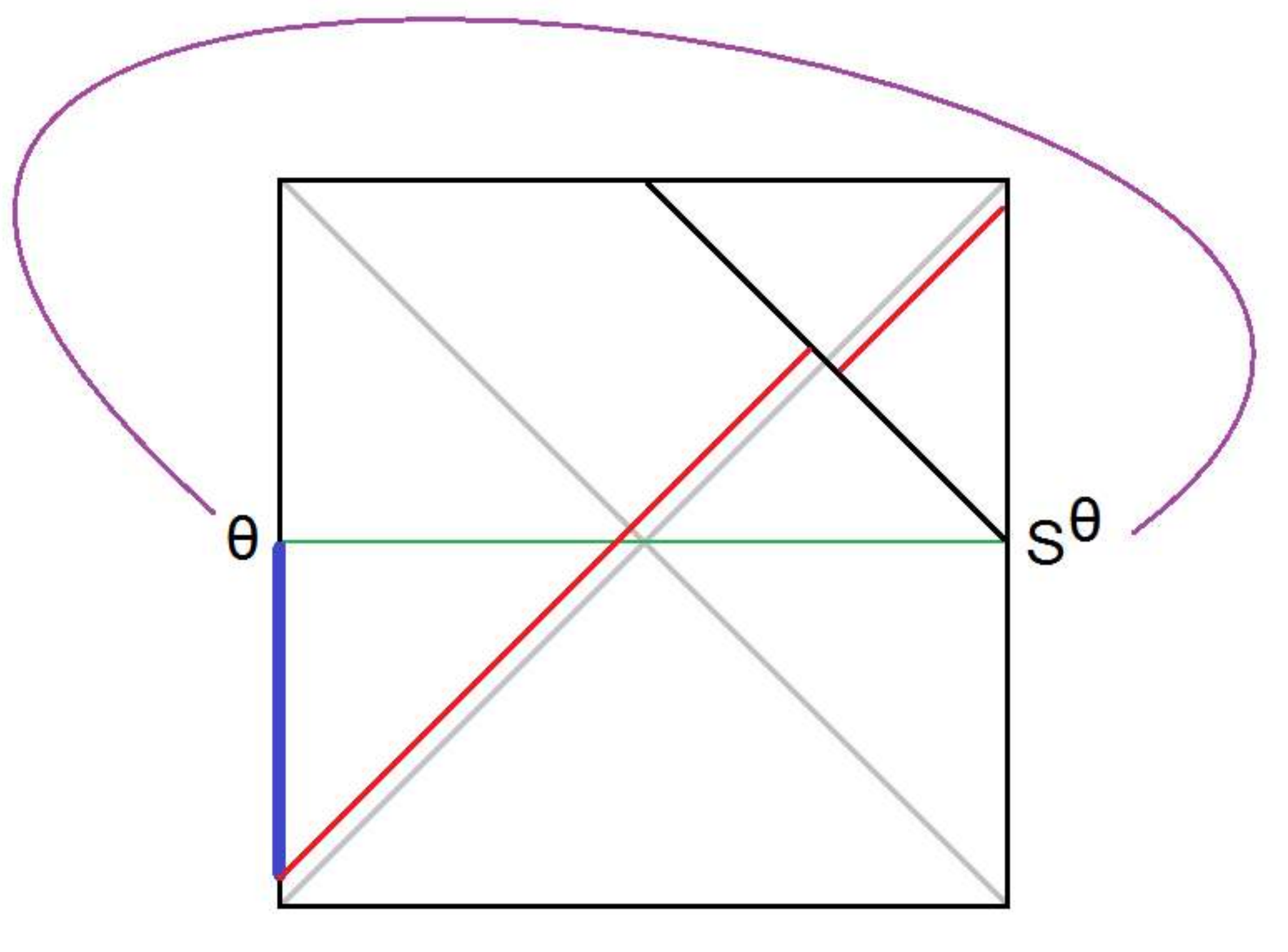}
\caption{Boosting the previous figure allows the coupling to act at $t=0.$ The picture now closely resembles the geometry described in \cite{Gao:2016bin}  and \cite{Maldacena:2017axo}.}
\label{Penrose7}
\end{center}
\end{figure}

\begin{figure}[H]
\begin{center}
\includegraphics[scale=.4]{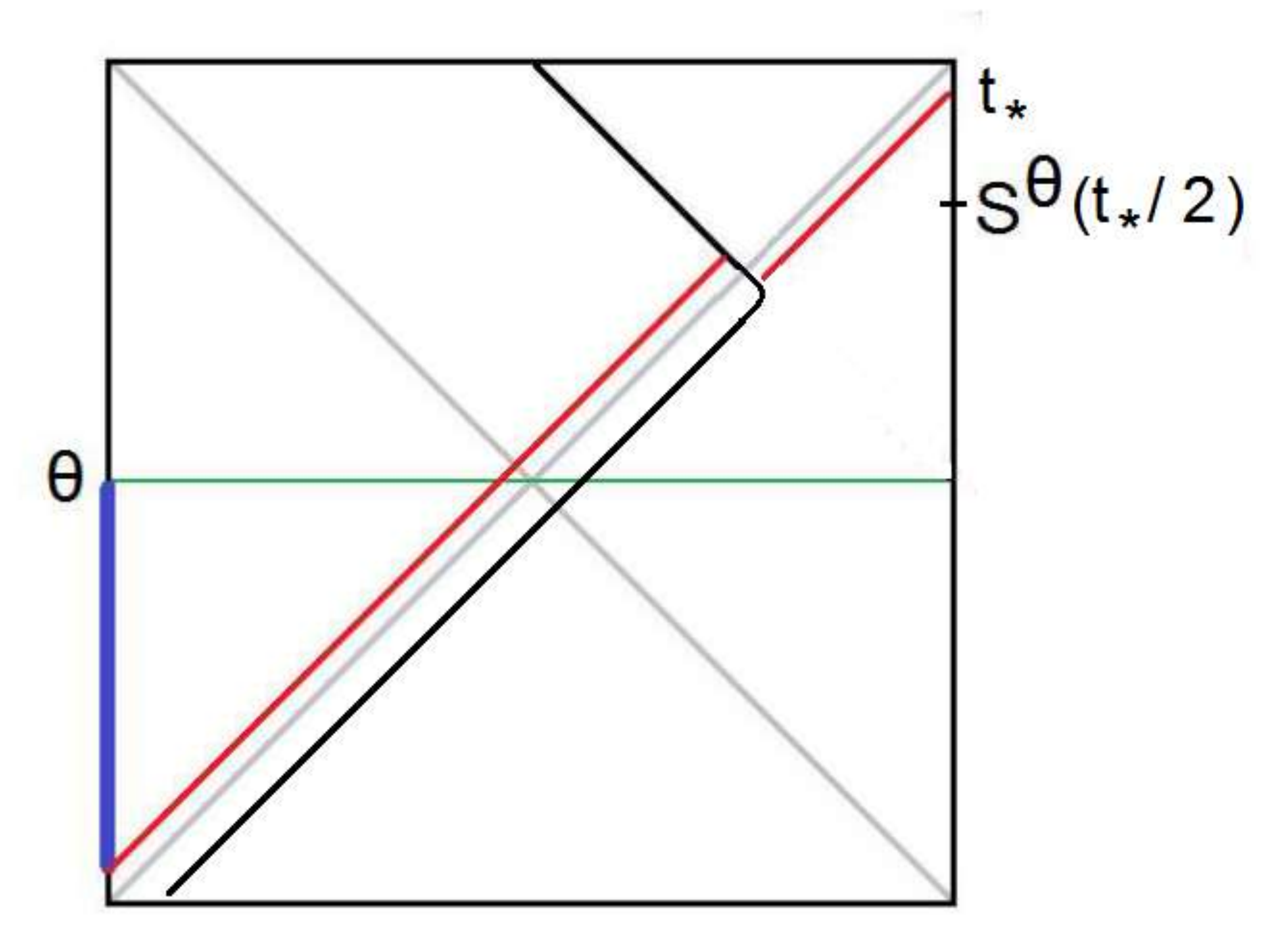}
\caption{This figure is the same as figure \ref{Penrose7} except that the shockwave turns back toward the horizon at time $t= -t_*/2.$}
\label{kink-shock}
\end{center}
\end{figure}

With this pretense we can see the similarity to \cite{Gao:2016bin}\cite{Maldacena:2017axo}. Let us boost the diagram by the usual boost symmetry so that 
the teleportee enters the geometry at $t= - t_*$ and Alice's measurement is at $t=0$. At the same time we must shift Bob's time so that the fictitious unit complexity operator acts at $t=0$ and the teleportee appears on Bob's side at $t=t_*.$ This is shown in figure \ref{Penrose7}.
Figure \ref{Penrose7} is essentially the same as the corresponding diagram for the protocol in \cite{Gao:2016bin}\cite{Maldacena:2017axo}. We conclude that the gravitational dual of our protocol must also be similar. Nevertheless we should keep in mind that                             treating the complexity of  $S^{\theta}(t_*)$ as being of order unity is a fiction.
The correct description involved the shock wave turing back toward the past horizon when it reaches the point $t=-t_*/2.$ As it happens this occurs just before the teleportee crosses the shock wave and comes out on Bob's side. This is shown in figure \ref{kink-shock}.

\bn
The shock wave makes its closest approach to Bob's boundary at $t_*/2$ and then, as it proceeds to the past, it turns back toward the past horizon.

One thing we notice from figure \ref{kink-shock} is that it is no longer necessary  for Alice's message to travel backward in time.

\section*{\small A Limitation of the Protocol}

By a successful teleportation of a system of $n$ qubits we mean that a subsystem of Bob's qubits materializes, unentangled with all other qubits, in a state identical to the initial state of $\bf T$. If this occurs then by the no-cloning principle, it is not possible for a copy of $\bf T$ to remain with  Alice.

In the present setup $\bf T$ would not escape from Bob's black hole; it would merely re-scramble and fall back in. In order to for $\bf T$ to escape we need to add more frozen qubits on Bob's side. By a swap operation the information in $\bf T$ can be transferred to the frozen qubits and escape among the higher energy non-thermal degrees of freedom. This last step is built in to the protocols of 
\cite{Gao:2016bin}\cite{Maldacena:2017axo}.

\section{Comparison of Protocols}

Our  circuit protocol for  quantum teleportation is similar to, but not quite the same as, 
 the traversability mechanism described in \cite{Gao:2016bin}\cite{Maldacena:2017axo}. In this section we will compare the two.
 
 Let us go back to figure \ref{Penrose7} keeping in mind that it involved a fiction; namely that the complexity of $S^{\theta} (t)$
 could be extrapolated back to order unity as $t$ approached $t_*.$ 
   The main difference between the protocols is that figure \ref{Penrose7} is actually correct for the protocols  of \cite{Gao:2016bin}\cite{Maldacena:2017axo}. Going back to equation \ref{U}, let us set $t=t_*.$ The Gao-Jafferis-Wall protocol is defined by setting the operator $\CU {(t_*)}$
to
\be 
\CU(t_*) = \exp{\{i {\CO}_L (t_*) }{\CO}_R (t_*)\}
\label{GJW}
\ee
where ${\CO}_{L,R}$  is a thermally smeared hermitian local CFT operator in the  left (right) CFT\footnote{In \cite{Gao:2016bin}  a small coupling constant $g$ appears in the exponent of \ref{GJW}. However, in order to reliably teleport a single qubit the coupling must be of order $1$. The strategy in \cite{Maldacena:2017axo} is to replace the single term in the exponent of \ref{GJW} by a sum over many left-right pairs of operators. This improves the reliability of the teleportation without increasing the basic coupling constant. 

In our protocol the reliability is also not perfect because the unitarity of the matrix ${V^{\dag}}^{\theta,\gamma}_{I,j}$ is not exact. It can also be improved by increasing the size of the subsystem $\theta.$  We thank Patrick Hayden for pointing this out.}.
  
If we assume that the spectrum of ${\CO}_{L}$ consists of the numbers $\theta$  (possibly continuous)  we may write 
\be 
\CU(t_*) =\sum_{\theta} \Pi_{\theta}(t_*) \ \exp{\{i \theta }{\CO}_R (t_*)\}.
\label{Uwall}
\ee 

Equation \ref{Uwall} has the same form as \ref{U} but with 
\be 
 S^{\theta}(t_*)=  \exp{\{i \theta }{\CO}_R (t_*)\}
 \ee
 
 Since $\CO$ is a thermally smeared operator we expect $S^{\theta}(t_*)$ to be a low complexity operator with complexity of order $1$. 
 
On the other hand  the precursor trick allows us to operate on Bob's side at $t=0$ with the precursor operator
 \be 
  S^{\theta}(0)=      V^{\dag} S^{\theta}(t_*)  V
 \ee
  Since $S^{\theta}(t_*)$ is a simple local operator the precursor $  S^{\theta}(0)$ has complexity $N,$ which agrees with the circuit protocol
in which the complexity of $ S^{\theta}(0)$ is also $N$. 

The operator $\CU$ in \ref{GJW} is symmetric with respect to the left (Alice) and right (Bob) sides. It can be used both to teleport a qubit from Alice to Bob and from Bob to Alice. The corresponding operator in \ref{U} was engineered for the purpose of teleporting from Alice to Bob and it is not apparent that that it has any symmetry that would enable teleportation from Bob to Alice.
  
\section{What Does Tom See?}

Let's suppose that the teleportee is a sentient being named Tom. We would like to know what Tom sees as he passes between Alice and Bob. Even more important, are the things Tom encounters recorded in his memory so that he can report them to Bob? At first sight it would seem not to be the case; the protocol was engineered so that Tom exits the wormhole in the same state as he entered.  However the protocol was constructed assuming the initial state of the  mediator was the TFD which does not contain anything interesting for Tom to encounter. To address the question we should vary the initial state away from the TFD, while keeping the rest of the protocol unchanged. 
Variations in the initial state will certainly affect  Tom's final state, but we have found it difficult to analyze directly in the circuit model. On the other hand the bulk dual offers an easier way to think about the problem of Tom's experiences. Let's consider an example in which Tom encounters a photon coming out of Alice's black hole  as he himself enters it. Let's go back to figure 
\ref{Penrose6} and modify the state at $t=0$ to include the photon. Let's suppose the photon would reach the Alice's boundary at time $t'$ if Tom where not there to intercept it. We can represent such a photon by a boundary operator $W_{AT}(t')$  (The subscript $A$ indicating an operator acting on the $\CA\CT$  system).  The initial TFD state is replaced by
\be 
|TFD'\ra = W_{AT}(t')|TFD\ra
\ee 

\begin{figure}[H]
\begin{center}
\includegraphics[scale=.4]{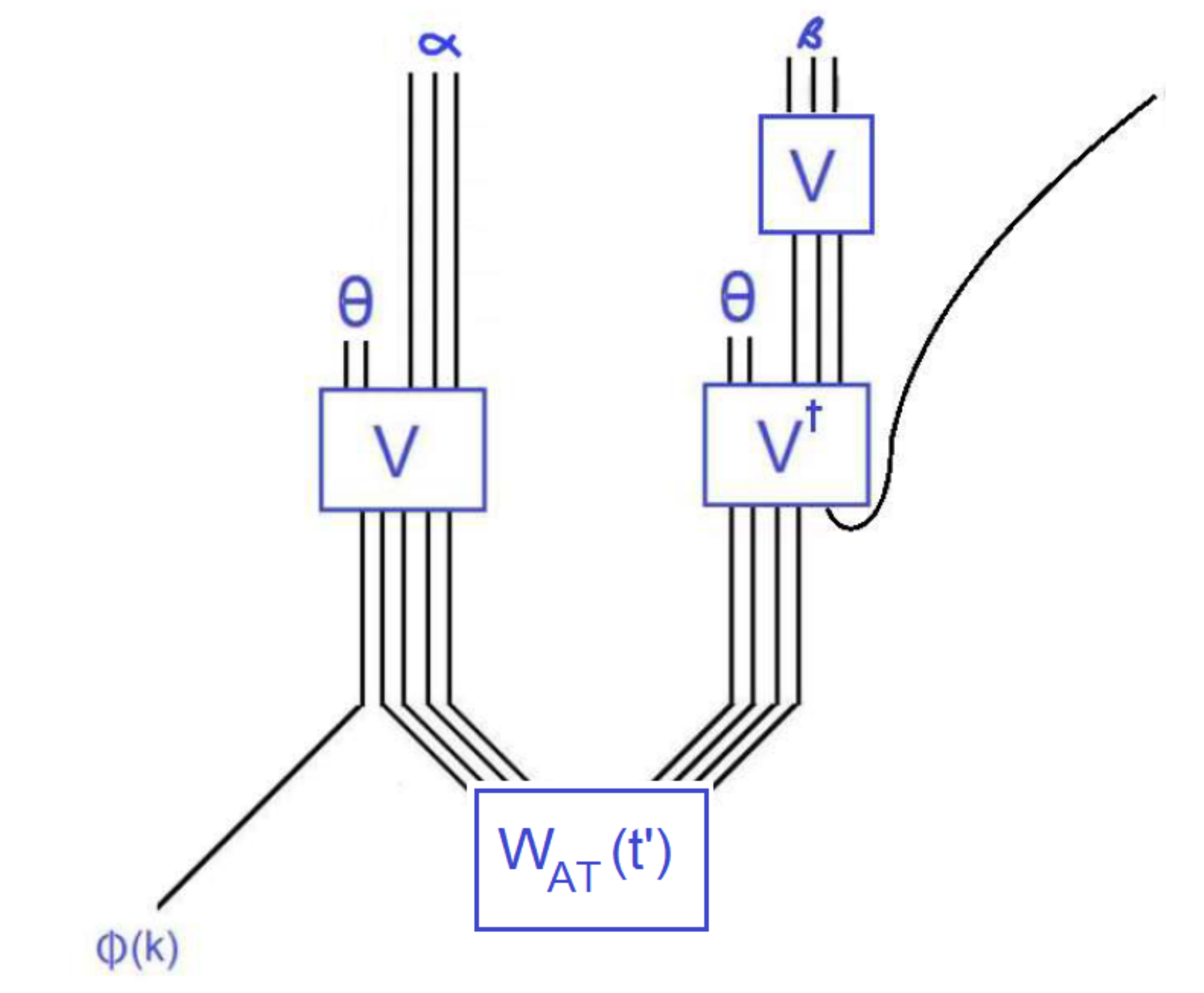}
\caption{The initial TFD state is modified by the insertion of a photon.}
\label{photon-insert}
\end{center}
\end{figure}

\begin{figure}[H]
\begin{center}
\includegraphics[scale=.75]{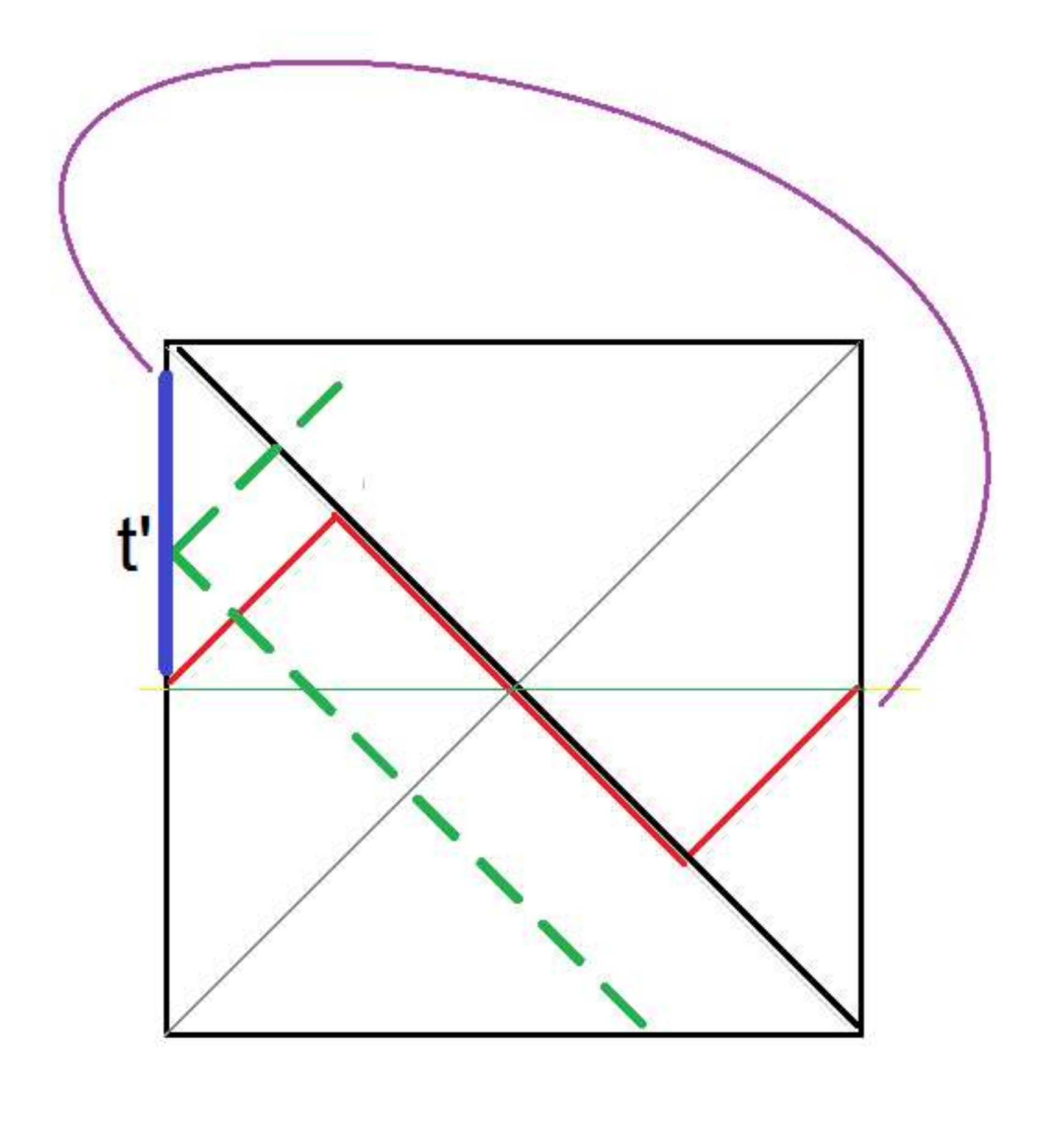}
\caption{This figure is the same as figure \ref{Penrose6} with the addition of an extra photon added to the TFD state. The extra photon is shown as a broken green line.}
\label{sentient}
\end{center}
\end{figure}

We now carry out the original protocol, replacing the initial entangled mediator state by $|TFD'\ra$. The circuit diagram would look like figure \ref{photon-insert}
\bn
which in itself is not very illuminating. The Penrose diagram in figure \ref{sentient} for the protocol is more interesting
It shows Tom (red trajectory) colliding with the photon (broken green trajectory) before hitting the negative energy shock wave. The collision with the photon will have various effects that we expect to show up in Tom's final state.

\section{ ER=EPR in the Lab}

The operations we've described would be very hard to do, if not impossible, for real black holes, but we can imagine laboratory settings where similar things may be possible. Suppose that in our lab we have two  non-interacting large  shells of matter, each of which has been engineered by condensed matter physicists to support conformal field theories of the kind that admit gravitational duals. If we make the two shells out of entangled matter we can produce the  shells in the thermofield double state for some temperature above the Hawking-Page transition. 

There is no obstruction to constructing the shells so that the speed of signal  propagation  in the shells is much slower than the speed of light in the laboratory. The experimenters would  not be limited by the signal velocity in the shells, and could run back and forth between the shells in a time which is negligible from the CFT perspective. We may also suppose that without violating any laws of quantum mechanics they could measure any observable, and apply any unitary operator to either shell. 

We can also eliminate any gravitational restrictions---for example concerns  that the shells would gravitationally collapse---by assuming that the gravitational coupling is as small as necessary. To summarize, we may assume the laboratory system is non-relativistic and that gravity exists but is negligible. 

Would there really be a hidden wormhole connecting the  shells? Assuming that the entire laboratory is embedded in a world that satisfies ER=EPR, then yes,  there would be such a wormhole. To confirm this
 Alice and Bob can merge themselves with the matter forming the shells and eventually be scrambled into the CFT thermal state. In the dual gravitational picture they would each fall into their respective black holes, and if conditions were right, they would  meet before being destroyed at the singlularity; all of this taking place in some space outside ordinary spacetime.  Unfortunately they would not be able to  inform the exterior world that the wormhole is real or that they successfully met.

However,  quantum teleportation  allows Alice and Bob to confirm the existence of the wormhole without jumping in. Alice may convince Tom to jump into the  wormhole, which she and Bob can render traversable by introducing a temporary coupling ( called  $\CU$ in this paper).
When Tom emerges out of the \dof \ of Bob's shell, he will recall everything he encountered, and can confirm that he really did traverse the wormhole. 

More practical than physical shells, two entangled quantum computers can simulate the CFTs. Of course to allow the teleportation of a real sentient being, the numbers of qubits in each computer would have to be enormous, but with a pair of hundred-qubit computers  a ten-qubit teleportee could be teleported. By allowing small variations of the initial state it should be possible to confirm  that the teleportee's final state responds to conditions in the wormhole, thereby giving  operational significance to ER=EPR.

\bn

We have made a very radical leap  in claiming that an  Einstein-Rosen geometry exists, connecting the two entangled shells or quantum computers, even though there are no real black holes in the lab. On the face of it this seems somewhat fantastical, but given that the lab is part of a quantum-gravitational world in which ER=EPR, the conclusion seems inevitable.

There is one final point: Carrying out such an experiment requires overcoming a large complexity obstacle. Injecting Tom into the AdS black hole is not simple. If we go back to figure  \ref{Penrose7} we see that Tom must enter the geometry in the remote past. This either requires that we prepare the system in a highly complex state of decreasing complexity---a white hole---or inject Tom as a complex non-local  precursor at $t=0.$ Fortunately, because of the switchback effect the complexity of the precursor is only of order $N$ but that's still complex. In particular the measure of complexity  is the entropy of the black hole and not the smaller number of degrees of freedom of Tom. This seems to be true of both  the protocols in this paper and those of \cite{Gao:2016bin}\cite{Maldacena:2017axo}.
Nevertheless, there does not seem to be an in-principle obstruction to \it laboratory teleportation through the wormhole. \rm A discussion of complexity of teleportation in the lab is given in Appendix \ref{App: labcomplexity}.

\bn

\section*{Acknowledgements}

We thank Ahmed Almheiri, Adam Brown, Hrant Gharibyan, Patrick Hayden, Daniel Jafferis, Juan Maldacena, and Douglas Stanford for discussions. 

Support for the research of L.S. came through NSF Award Number 1316699. 

\sc
\appendix

\section{Large System Teleportation}\label{App: Large}

The generalization to larger teleportees consisting of $n$ qubits is straightforward as long as $n<<N.$ The circuit diagrams \ref{small-teleportation}, \ref{crossing}, \ref{zee}, \ref{tR=0}, \ref{subfinal}, \ref{grid1}, \ref{grid2}, \ref{grid3} are essentially unchanged, except that:
\bi 
\item  The single qubit $k$ is replaced by $n$ qubits. The state 
$|\Phi\ra $  is now an $n$ qubit state.
\item Alice measures $2n$ qubits. The symbol 
symbol $\theta$ represents a bit-string of length $2n.$ 
\item  The symbols $|\alpha\ra$ and $  |\beta\ra$ represent states of $(N-m)$ qubits.
\item The number of classical bits that Alice sends to Bob is $2n.$
\item The reduction in entanglement entropy required to teleport the $n$-qubit state is $n$.
\ei

In \cite{Susskind:2014yaa} the limiting  case $n=N$  was studied, in which all three systems have $N$-qubits. The teleportee is as large as each member of the mediator. 
  At the end of the teleportation all the entanglement will have been used up. We can call it ``large system teleportation". A special case would be the teleportation of a black hole of entropy $S$ through an Einstein-Rosen bridge with entanglement entropy $S$.  Since the teleportation of a system of $n$ qubits must deplete the entanglement entropy of the mediator by at least $n$, large system teleportation represents the maximum amount of information that can be teleported by a given mediator.  In this appendix we will review large system teleportation   and discuss some subtleties.

Let us suppose that 
 the initial state of the mediator is  maximally entangled in the infinite temperature Thermofield-double (TFD) state, 
 \be
|TFD \ra = \sum_I |I\ra_A  \ |I\ra_B
\label{TFD2}
\ee
where $|I\ra_{A,B}$ indicate a complete set of states in the computational bases of Alice's and Bob's systems.

Alice is also be in possession of the teleportee, in this case another $N$ qubit system. The state of the teleportee is pure, and has the form
\be 
|\Phi\ra = \sum_{K}  \Phi(K) \  |K\ra_T
\ee
where $K$ is a complete set of computational states in the Hilbert space of the teleportee labeled $T$. The initial state is,
\be 
|initial\ra = \sum_{IK} \  \Phi(K) \  |K\ra_T  \ |I\ra_A  \ |I\ra_B.
\ee

\begin{figure}[H]
\begin{center}
\includegraphics[scale=.6]{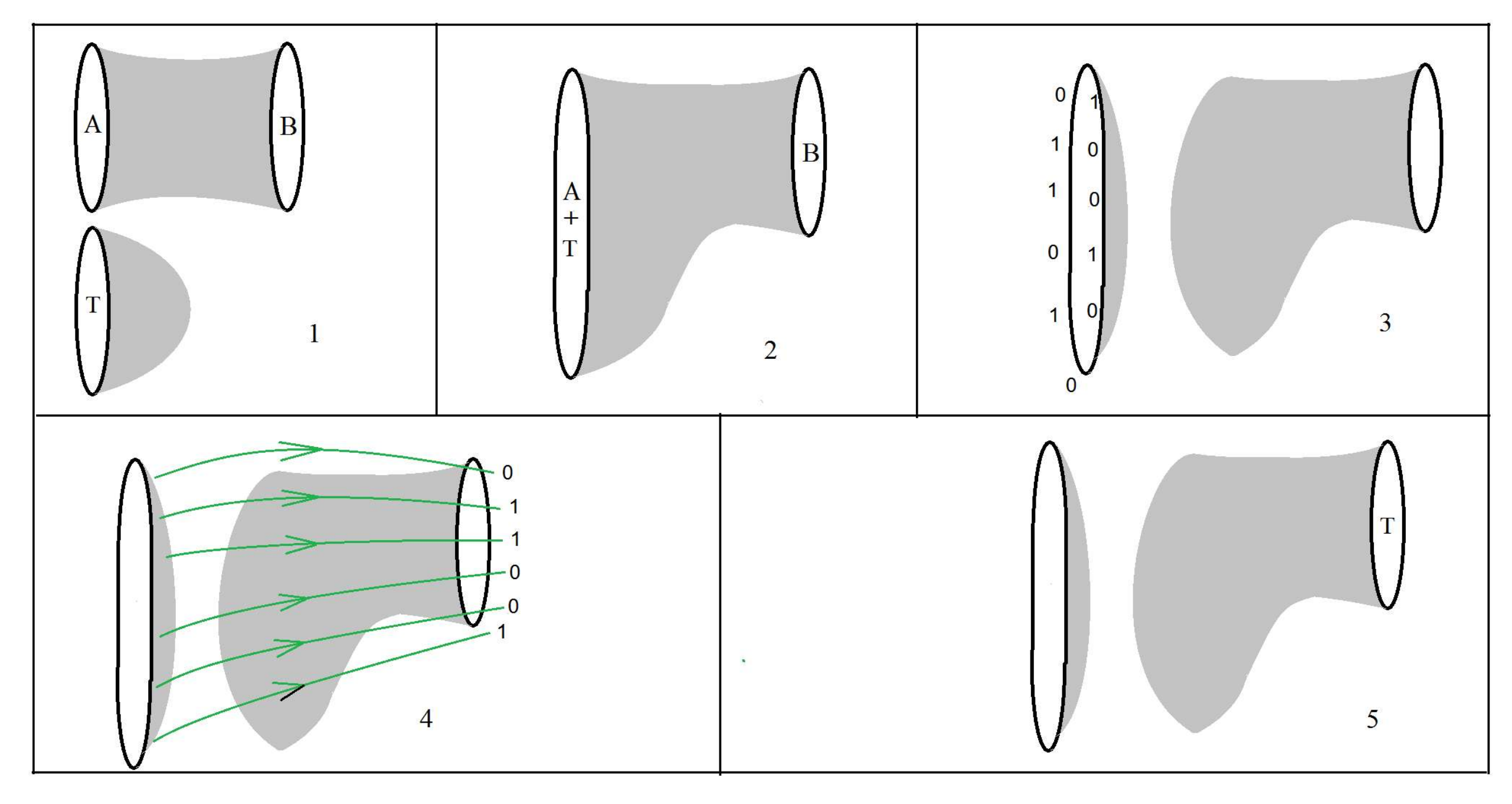}
\caption{The figures are adapted from the second reference in \cite{Susskind:2014yaa}. They illustrate the five steps described below: 
1) Alice combines her share of the mediator, $A,$ with the teleportee $T$:  2) The combined system $(\bf AT)$ evolves for a scrambling time:  3) Alice performs a complete set of measurements on $(\bf AT)$ in the computational basis:  4) Alice sends the classical outcome to Bob:  5) Bob applies a unitary rotation, thereby  transforming the state of his  black hole  to $|\Phi\ra$.}
\label{five-step}
\end{center}
\end{figure}

The goal is to teleport the state $|C\ra$ from Alice to Bob.
The protocol can be summarized by five steps illustrated in the cartoon strip in  figure \ref{five-step}.

\begin{enumerate}
\item Alice combines the systems $A$ and $T$ into a single composite system of $2N$ qubits called $(\bf AT)$. The initial state is re-written
\be 
|initial\ra = \sum_{IK}  \Phi(K) \ |KI\ra_{AT} \  |I\ra_B.
\ee

\item Alice allows    the composite $2N$-qubit system $(\bf AT)$ to evolve  until it becomes  scrambled  \cite{Sekino:2008he}. This takes time $t_* = N \log N.$  For the case of black holes this means that  $A$ and $T$ merge into a single black hole in  equilibrium. 

 Let the unitary scrambling operator on $(\bf AT)$ system be called $V$,
 $$V = e^{-iHt_*}.$$
  The state evolves to
\be 
V|initial\ra = \sum_{IK}  \Phi(K) \  V|KI\ra_{AT} \  |I\ra_B.
\ee

\item Alice measures all $2N$ qubits of the $(\bf AT)$ system in the computational basis. Let the outcome be the string of binary digits $\theta,$ and the let the corresponding computational state be $|\theta\ra.$ Also let the projection operator onto $|\theta\ra$ be $\Pi_{\theta} = |\theta\ra \la \theta|.$  (Note that $|\theta\ra$ is a state in the $\bf AT$ system and that $\Pi_{\theta}$ acts in the $(\bf AT)$ factor of the full Hilbert space.)

The measurement collapses the state to,

\be 
 \Pi_{\theta} V|initial\ra = \sum_{IK}  \Phi(K) \  \la \theta| V|KI\ra_{AT} \  |I\ra_B \ |\theta \ra .
 \label{collapse}
\ee
  \bn
  
 Consider the matrix $\la \theta | V|KI\ra = V^{\theta}_{KI}$ for fixed outcome $\theta.$ In general it has no special property  but because $V$ is a scrambler, for any given $\theta$ the matrix $V^{\theta}_{KI}$ is unitary,
\be 
\sum_I \ \la \theta| V|KI\ra  \  \la I L | V^{\dag} |\theta \ra = \delta_{KL}
\ee
The unitarity of  $V^{\theta}_{KI}$ is an important simplification.
It follows that the Bob-state in \ref{collapse},
 \be 
 \sum_{IK}  \Phi(K) \  \la \theta| V|KI\ra \  |I\ra_B \ .
 \label{unitary}
\ee
is unitarily related to 
$$
\sum_K \Phi(K) |K\ra_B,
$$
but with the unitary operator $V^{\theta}$ being dependent on $\theta.$

\item
Alice sends the outcome string $\theta$ to Bob in the form of $N$ classical bits.

\item
When Bob receives the classical message he knows what operator  to apply in order to undo the unitary matrix in \ref{unitary}. That operator is ${V^{\theta}}^{\dag}$ acting on Bob's factor of the Hilbert space.

If the protocol is carried out the final state is given by,
\be 
|final \ra =  |\Phi\ra_B   
|\theta\ra_{AT}.
\ee
The teleportation has been accomplished. 
Meanwhile on Alice's side the $2N$ qubits are left in the state 
$
|\theta\ra_{AT} 
$.

\end{enumerate}

To make clear the simplification that the scrambling introduces,  recall that in the simplest quantum teleportation of a single qubit, Alice is required to measure the $(\bf AT)$ system in the Bell basis, not the computational basis. If she wants to measure in the computational basis she must first  apply a unitary operator to rotate from the Bell basis to the computational basis. Similar things are true for  large system  teleportation. In that case the scrambling operator plays the role of the rotation to the computational basis. 

\bn
In order to compare with \cite{Gao:2016bin} and \cite{Maldacena:2017axo}  the teleportation protocol can be slightly modified. The measurement operation, the classical communication, and the final unitary operation by Bob, can be combined into a single unitary operation that acts on both sides.
\be  
\CU = \sum_{\theta}\Pi_{\theta} \otimes {V^{\theta}}^{\dag}
\label{operator}
\ee

The operator $\CU$ projects the  $(\bf AT)$ state onto $|\theta\ra$;  then applies ${V^{\theta}}^{\dag}$ to Bob's side; and finally sums over $\theta.$
  Acting with the non-local operator $\CU$ is 
 the analog of the two-sided interaction that leads to traversability in \cite{Gao:2016bin}\cite{Maldacena:2017axo}.
 Instead of the final state of the Alice-Teleportee system being $|\theta\ra$ as in ordinary teleportation, acting with $\CU$ leaves $(\bf AT)$ in a linear superposition of  all $|\theta\ra,$

\be 
|final \ra =  |\Phi\ra_B  \otimes \sum_{\theta} |\theta\ra_{AT} 
\ee
\bn

\bn

Here is a circuit diagram representing the protocol. 

\begin{figure}[H]
\begin{center}
\includegraphics[scale=.3]{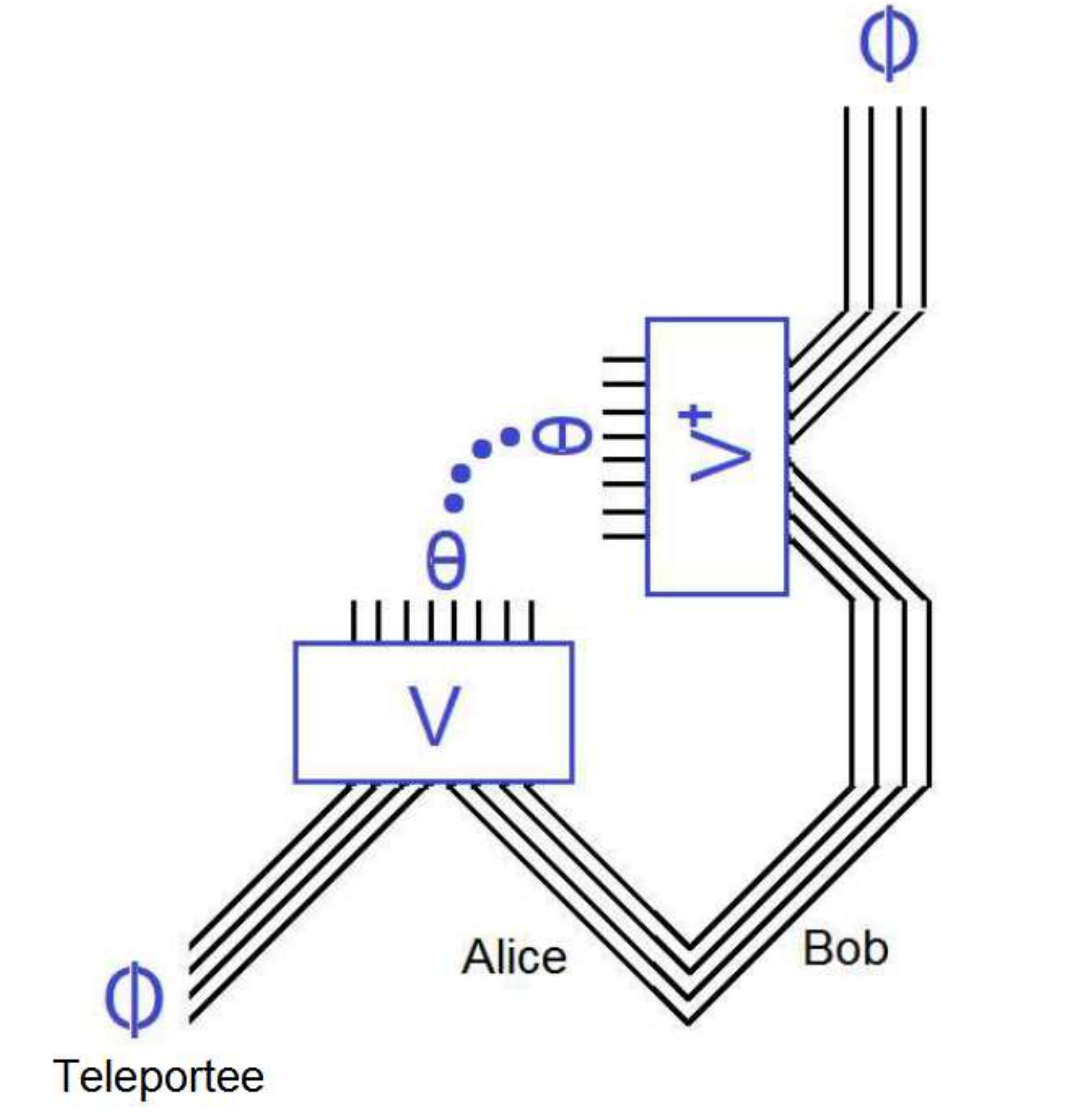}
\caption{Circuit diagram for large system teleportation.}
\label{circuit1}
\end{center}
\end{figure}

Notice that on Alice's side $V$ is an operator that connects a state of $2N$ qubits to another state of $2N$ qubits, whereas on Bob's side $V^{\dag}$ is thought of as an operator, parameterized by $\theta$, and acting on an $N$ qubit state to give another $N$ qubit state.

\bn

\section*{\small Frozen Qubits and the Teleportion of Black Holes}

There is a subtlety which would prevent the protocol of figure \ref{circuit1} from being applied to teleporting black holes without some modification. The modification  is required because the merging and scrambling of black holes is not an adiabatic process; in fact it generates a significant amount of thermodynamic entropy. We made an unrealistic assumption, namely that the number of qubits is conserved during the merging of Alice's share of the mediator and the teleportee. For example suppose all three systems  are $(3+1)$ dimensional Schwarzschild black holes of mass $M,$ and entropy,
\be 
S= 4\pi M^2 G = N  \log{2}.
\ee

\begin{figure}[H]
\begin{center}
\includegraphics[scale=.4]{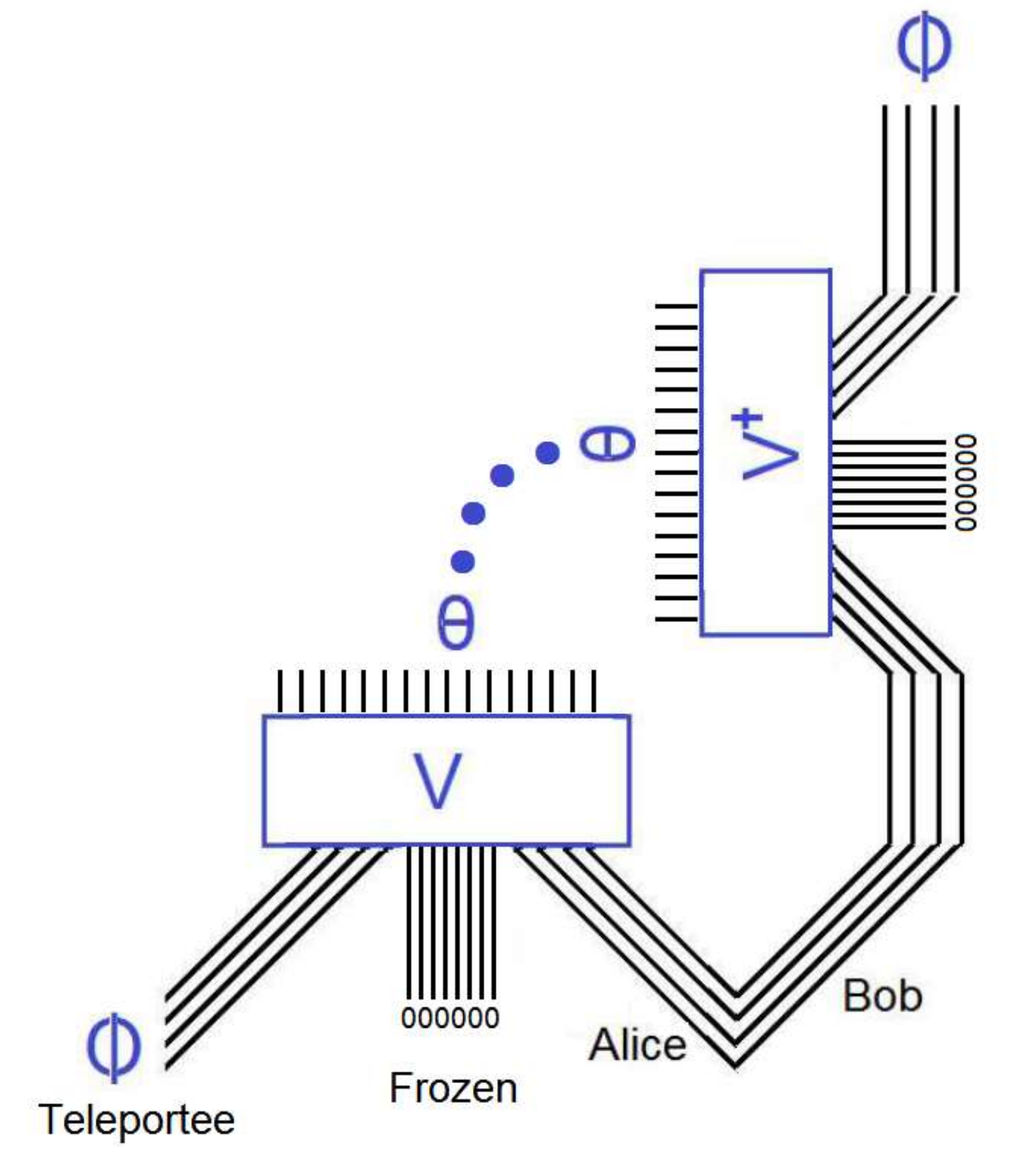}
\caption{Circuit diagram for the teleportation of a black hole.}
\label{circuit1-1}
\end{center}
\end{figure}

After the merger of $\bf A$ and $\bf T$ the resulting black hole $\bf (AT)$ will have mass $2M$ and entropy $4S$. The required number of qubits to describe the system  is $4N$.  
To describe this situation we may assume that the $(\bf AT)$ system is described by $4N$ qubits of which $2N$ were initially excited to form the $(\bf AT)$ system,  and the rest were ``frozen" into the state $|0\ra^{\otimes 2N}$. The scrambling operation will excite the frozen degrees of freedom by acting on all $4N$ qubits. The circuit diagram replacing figure \ref{circuit1} is shown in figure \ref{circuit1-1}
For fixed $\theta$ the matrix $V^{\theta}_{KI}$ is still unitary, but  the bit-string $\theta$ is of length $4N.$ The non-adiabatic merging requires that the teleportation protocol  transfers   $4N$ (rather than $2N)$ classical bits from Alice to Bob\footnote{In making this claim we assume that the measurement is done in the computational basis.}.

The existence of frozen qubits is natural in AdS/CFT where the number of degrees of freedom is infinite, but only a finite number are excited in a black hole configuration. The remainder are UV \dof \ which are inactive in the initial state but some of which become excited by the merger. The excitation of the frozen \dof \ increases the thermal entropy on Alice's side but leaves  the entanglement entropy unchanged.

\section{Complexity}\label{size and complexity}

Throughout this paper we have referred to the complexity of various precursor operators without explicitly defining the concept. In the context of this paper we can be precise by relating the complexity to the \it size \rm  \cite{Size} which appears in recent work on scrambling and chaos.

For simplicity let the simple operator $W$ be a one-qubit operator acting on the first qubit. Let the remaining qubit operators be labeled $Z_i.$ The precursor $U^{\dag}(t) W U(t)$ is denoted $W(t).$
With those conventions the size of the precursor is defined by,
\be 
s(t) = \sum_i \bf Tr \rm [ W(t), X_i]^2
\ee
where the boldface symbol $\bf Tr$ means normalized trace, i.e., $\bf Tr \it  I = 1$.

The size of a precursor grows exponentially with time until it saturates at the scrambling time,
\bea  
s(t)\eq e^t   \ \ \ \ \ t \leq t_*  \cr 
s(t) \eq N  \ \  \ \ \ t \geq t_*
\eea
The complexity of a precursor is related to the size by \cite{Size},
\be 
 \frac{\d\CC }{dt} = s(t).
\ee
It grows exponentially until the scrambling time and then increases linearly. In the exponentially growing region---the region of interest in this paper---the size and complexity are essentially the same.  Note that at the scrambling time $\CC \approx N.$

The fact that the complexity does not begin to grow linearly before the scrambling time is called the switchback effect.

\section{Complexity of decoding}\label{App: decodingcomplexity}

Before we look for the detailed protocol, let's estimate the decoding complexity. Here is what we mean by decoding complexity. Say, the message qubit $\Phi$ is in the decoder's density matrix $\rho$. That means, the decoder can apply an unitary operator $U$ ($U$ does not depend on $\Phi$), s.t., 
\begin{align*}
U\rho U^{\dagger} =|{\Phi}\ra\la{\Phi}|\otimes\rho'
\end{align*}
We ask for the minimal complexity of such $U$'s. 

To simplify, let's first consider non-traversable wormholes. Alice throws in the message qubit $\Phi$ when the black hole is in thermofield double, and waits for it to scramble. What's the complexity for her to decode the information? i.e, what's the minimal complexity of the operations she can do to separate out a qubit and put it in state $\Phi$?

\begin{figure}[H]
 \begin{center}
 \hspace{-2.5cm}
  \begin{subfigure}[b]{0.4\textwidth}
  \begin{center}
    \includegraphics[scale=0.45]{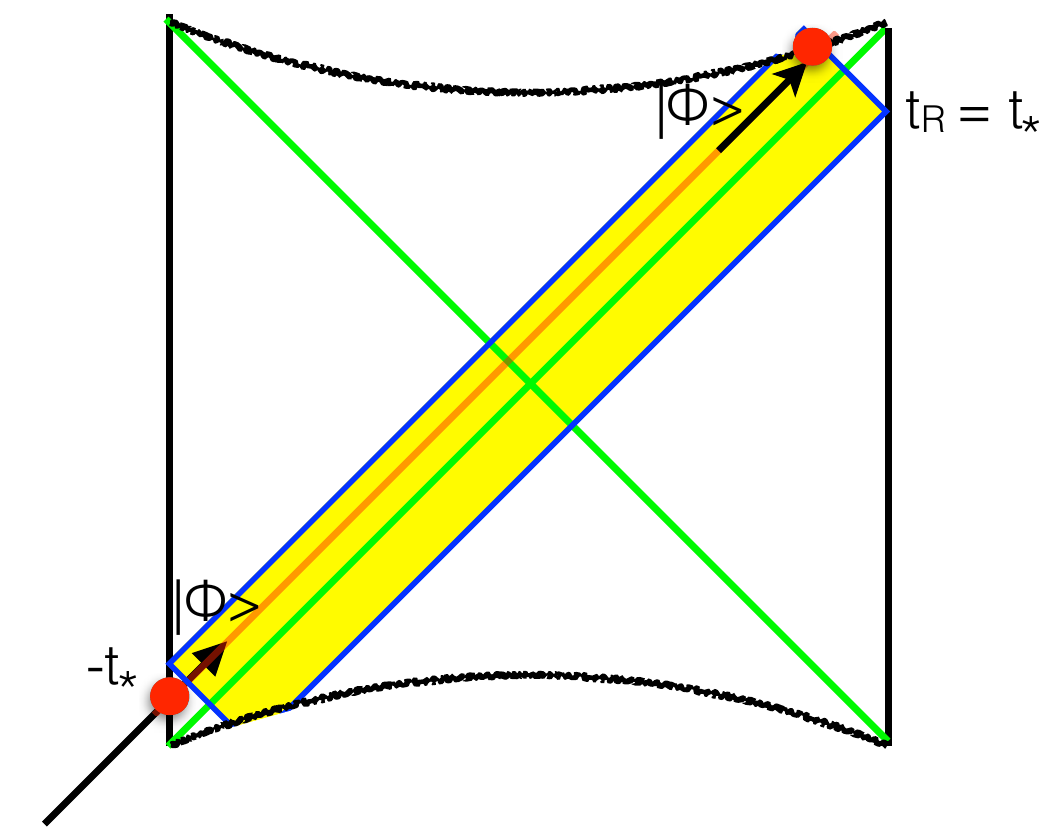}
    \caption{}
    \label{WDWnontraversable0}
    \end{center}
  \end{subfigure}
  \hspace{-1cm}
  \begin{subfigure}[b]{0.4\textwidth}
  \begin{center}
    \includegraphics[scale=0.45]{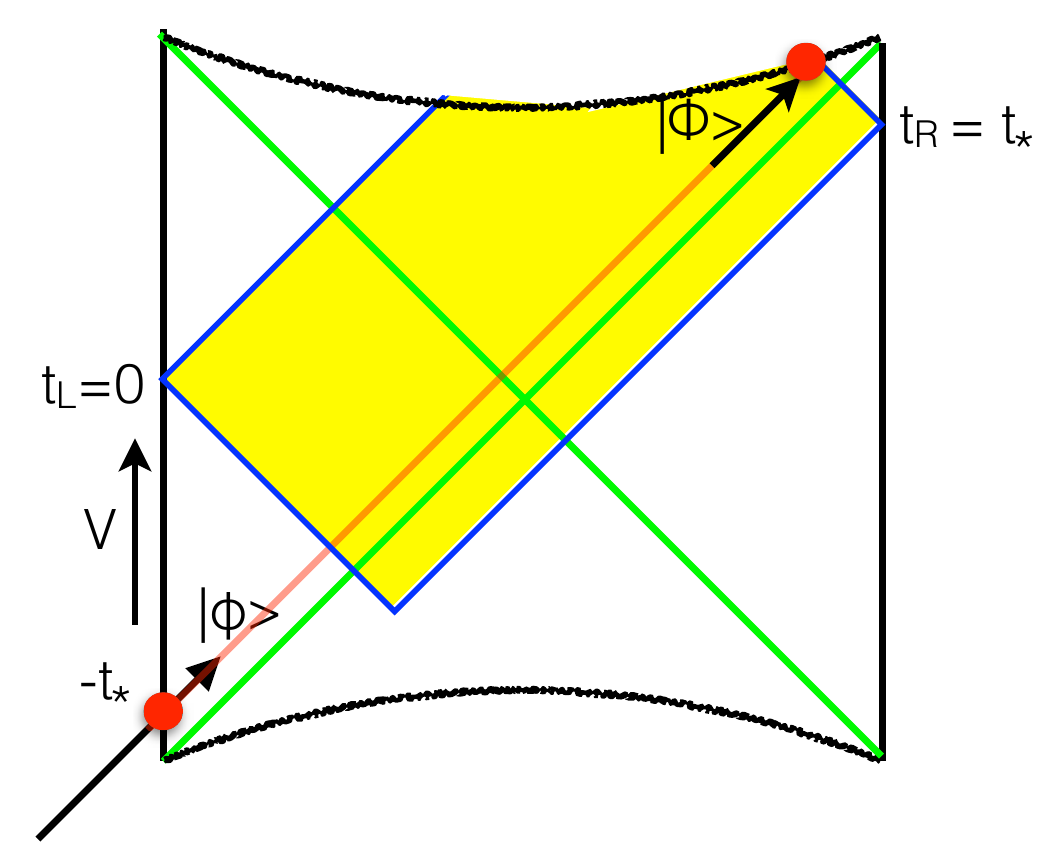}
    \caption{}
    \label{WDWnontraversable1}
    \end{center}
  \end{subfigure}
  \hspace{-1.8cm}
  \begin{subfigure}[b]{0.5\textwidth}
  \begin{center}
    \includegraphics[scale=0.45]{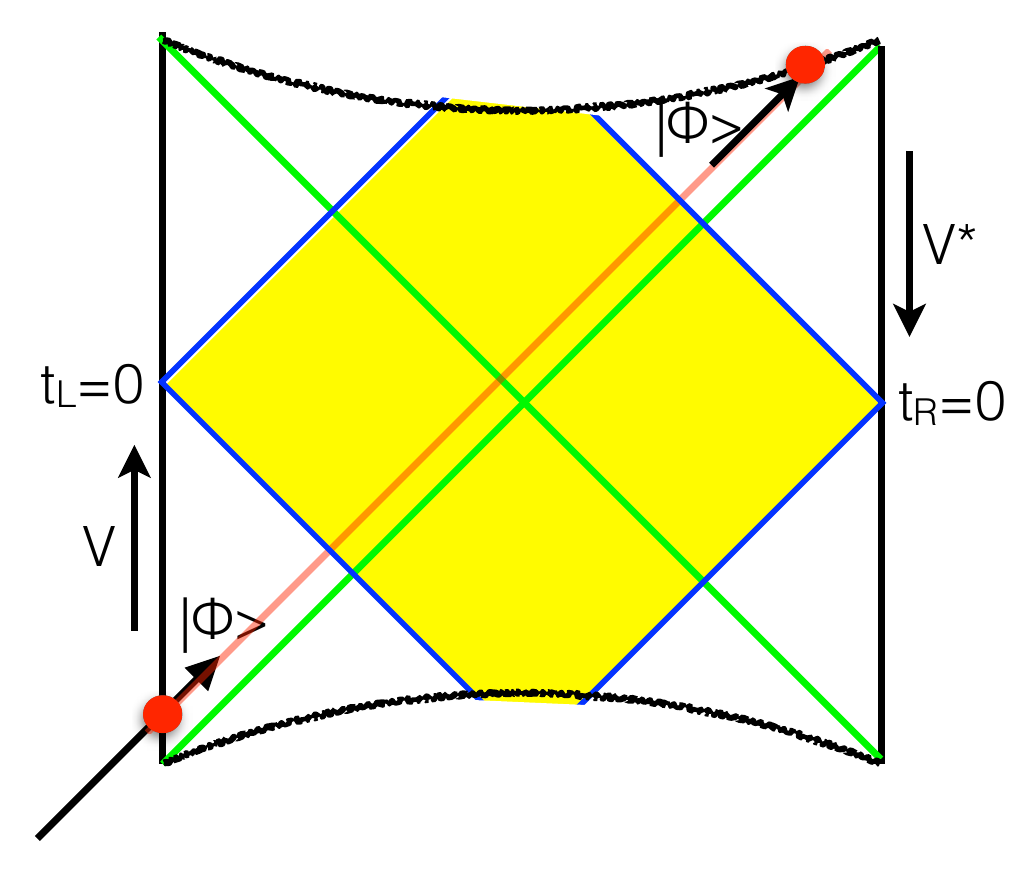}
    \caption{}
    \label{WDWnontraversable2}
    \end{center}
  \end{subfigure}
  \hspace{-3cm}
  \caption{Message in non-traversable wormhole}
  \label{WDWnontraversable}
  \end{center}
  \vspace{-.5cm}
\end{figure}

Say, we fix right time at $t_R = t_*$. The message is sent in by Alice at $t_L = -t_*$ (when the black hole is in thermofield double). (Figure \ref{WDWnontraversable0}) At this point her decoding complexity is $0$. 
\begin{align*}
\psi_0 = \sum_I |{\Phi,I}\ra_A|{I}\ra_B=\Phi(k)|{k,I}\ra_A|{I}\ra_B
\end{align*}
She waits for scrambling time, and looks at the state at $ t_{L} = 0, \ t_R = t_*$. (Figure \ref{WDWnontraversable1}) 
The state now becomes
\begin{align*}
\psi_1 = \Phi(k)V_{ iJ,kI}|{i, J}\ra_A|{I}\ra_B
\end{align*}
The relative complexity of $\psi_1$ and $\psi_0$ is scrambling complexity $N\log N$. Naively, to decode the message Alice needs to undo the scrambling and her decoding complexity is also $N\log N$. But in fact, it's not that large. Consider the epidemic picture \cite{Size}.  We throw in an extra qubit. At the first time step, the epidemic spreads to one other qubit. At the next time step, four qubits are affected. In the last time step, $\sim\frac{N}{2}$ qubits are affected. If we want to decode the message, the goal is to separate the extra qubit from others. We'll first clean the $\frac{N}{2}$ qubits who got affected at the last step, then the $\frac{N}{4}$ qubits affected at the last second step. So at the end of the day, we only need $\sim N$ gates to separate the epidemic qubit. The decoding complexity will be $\sim N$. After the decoding the rest of the qubits still stay pretty much scrambled. 

A more rigorous argument goes as follows. Alice' decoding procedure only depends on the left density matrix. In particular, a unitary transformation on the right side should not affect the decoding. We can instead look at the state at $t_L = 0, \ t_R = 0$. (Figure \ref{WDWnontraversable2}) 
\begin{align*}
\psi_2 =\ & \sum_{I,k,i, J,K}\Phi(k)V_{iJ, kI}|{i, J}\ra_AV^*_{K, I}|{K}\ra_B\\
=\ &\sum_{I,k,i, J,K}\Phi(k)V_{iJ, kI}V^{\dagger}_{I, K}|{i, J}\ra_A|{K}\ra_B
\end{align*}
Note that $\psi_2$ only has relative complexity $\sim N$ with $\psi_0$, since the operator $V_{iJ, kI}V^{\dagger}_{I, K}$ is a precursor at scrambling time. Alice can use inverse of that operator to decode the message. So the decoding complexity is $\sim N$. 

After the decoding, the state will become
\begin{align*}
\psi_3 = V^{(N)}V^{(N+1)\dagger}\psi_1 = \sum_{I, J}V_{J,I}|{\Phi, J}\ra_A|{I}\ra_B
\end{align*}

In fact, the above conclusion does not depend on two-sided black hole. What we learn from this is, to decode the message, one does not need to completely undo the scrambling. To minimize complexity, one can just separate the message qubit, while leaving the rest of the qubits scrambled. This only costs complexity $\sim N$. Say, we start from state $|{000...000}\ra$, add one extra qubit $|{\Phi}\ra$, and scramble it: $V|{\Phi, 000...000}\ra$. Here are the encoding and decoding circuits:

\begin{figure}[H]
 \begin{center}
  \begin{subfigure}[b]{0.4\textwidth}
  \begin{center}
    \includegraphics[scale=0.4]{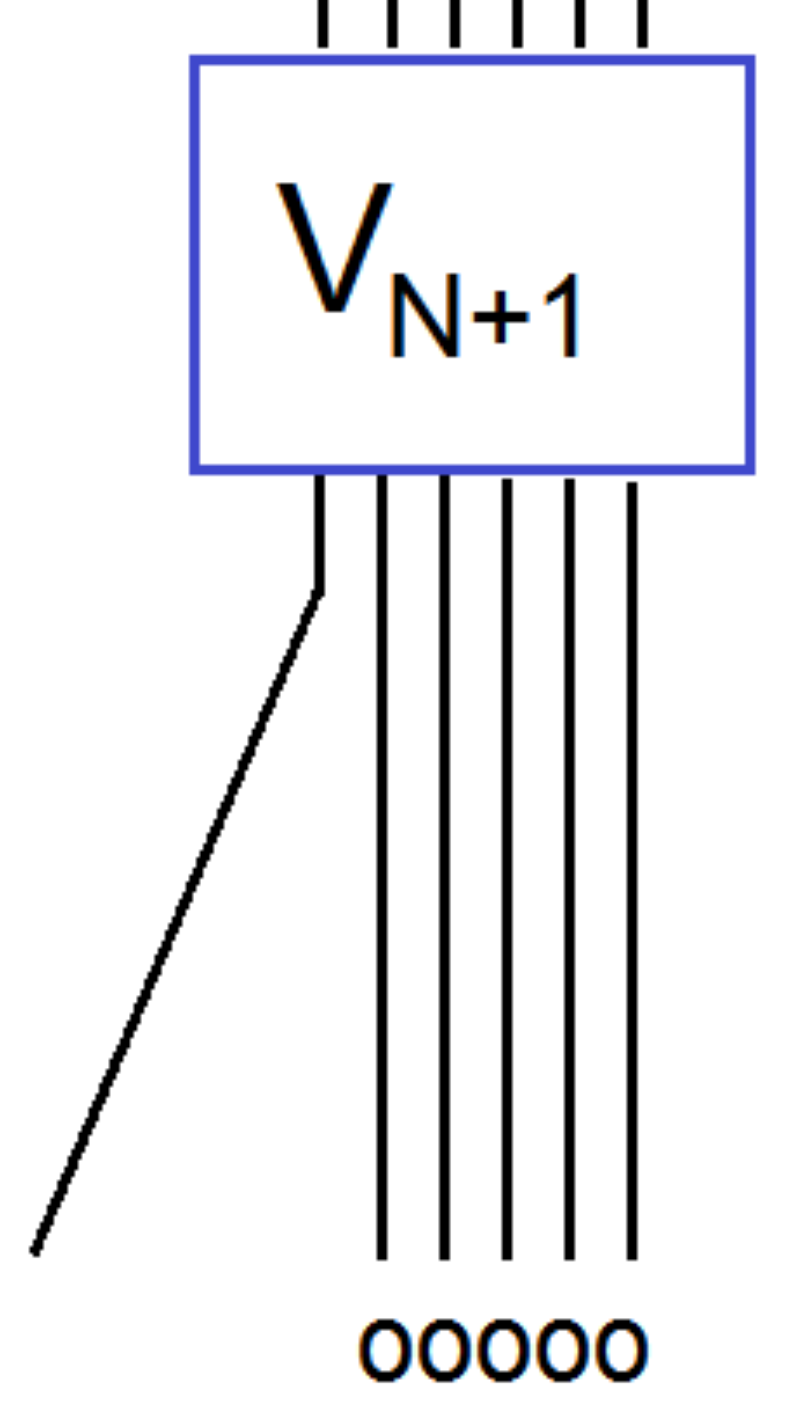}
    \caption{}
    \label{Encoding circuit}
    \end{center}
  \end{subfigure}
  \hspace{-1cm}
  \begin{subfigure}[b]{0.4\textwidth}
  \begin{center}
    \includegraphics[scale=0.4]{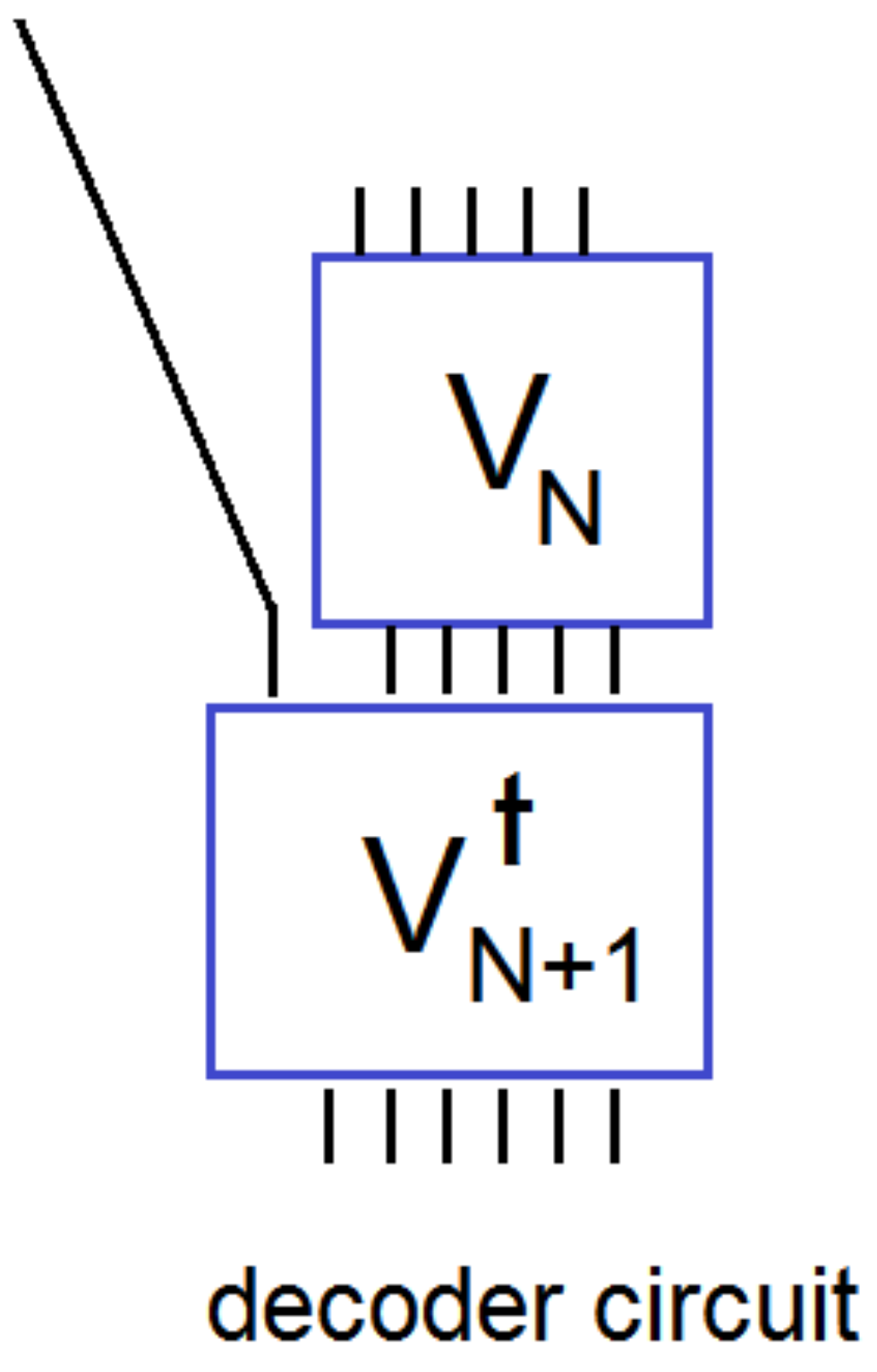}
    \caption{}
    \label{Decoding circuit}
    \end{center}
  \end{subfigure}
  \hspace{-1.8cm}
  \caption{Encoding and minimal complexity decoding}
  \label{encodingdecoding}
  \end{center}
  \vspace{-.5cm}
\end{figure}

Note that the decoding circuit (Figure \ref{Decoding circuit}) is a precursor and only has complexity $\sim N$. After the decoding, the state is $|{\Phi}\ra V|{000...000}\ra$. \\

Next, we consider traversable wormhole and Bob's decoding complexity. Say, Alice throws in the message, and waits for her system to get scrambled. Then she gives some qubits or classical bits to Bob. What's Bob's decoding complexity at this point?
Before she communicates with Bob, the state is
\begin{align*}
\psi_1 =\ & \Phi(k)|{i,J}\ra_AV_{ iJ,kI}|{I}\ra_B \\
=\ & \Phi(k)|{i,\theta,\alpha}\ra_AV_{ i\theta\alpha,kI}|{I}\ra_B
\end{align*}
where in the second line we write $|{J}\ra = |{\theta,\alpha}\ra$.

After Alice sends some qubits (or classical bits) $\theta$ to Bob, the state becomes
\begin{align*}
\tilde\psi_1 =\ & \Phi(k)|{i, \alpha}\ra_AV_{i \theta\alpha,kI}|{ \theta,I}\ra_B\\
\equiv\ &\Phi(k)|{ \theta,I}\ra_BU_{\theta I, ki\alpha}|{i, \alpha}\ra_A
\end{align*}

If we compare state $\psi_1$ and $\tilde \psi_1$, we see that they have quite similar structures. There is a scrambling matrix connecting two sides. To make the analogy more clear, in the second line of $\tilde \psi_2$ we write it in slightly different form. The only difference between $\psi_1$ and $\tilde \psi_1$ is that, in $\psi_1$ Alice has more than half of the total indices, while in $\tilde\psi_1$ Bob has more than half of total indices. By property of scrambling matrix, we know that in state $\tilde\psi_1$, the message is in Bob's hand. 

From earlier argument, we know that in state $\psi_1$, Alice' decoding complexity is $\sim N$, so we know that in state $\tilde\psi_1$, Bob's decoding complexity is also $\sim N$. 

What does this tell us about the dual bulk geometry?

\begin{figure}[H]
 \begin{center}
 \hspace{-2.5cm}
  \begin{subfigure}[b]{0.4\textwidth}
  \begin{center}
    \includegraphics[scale=0.5]{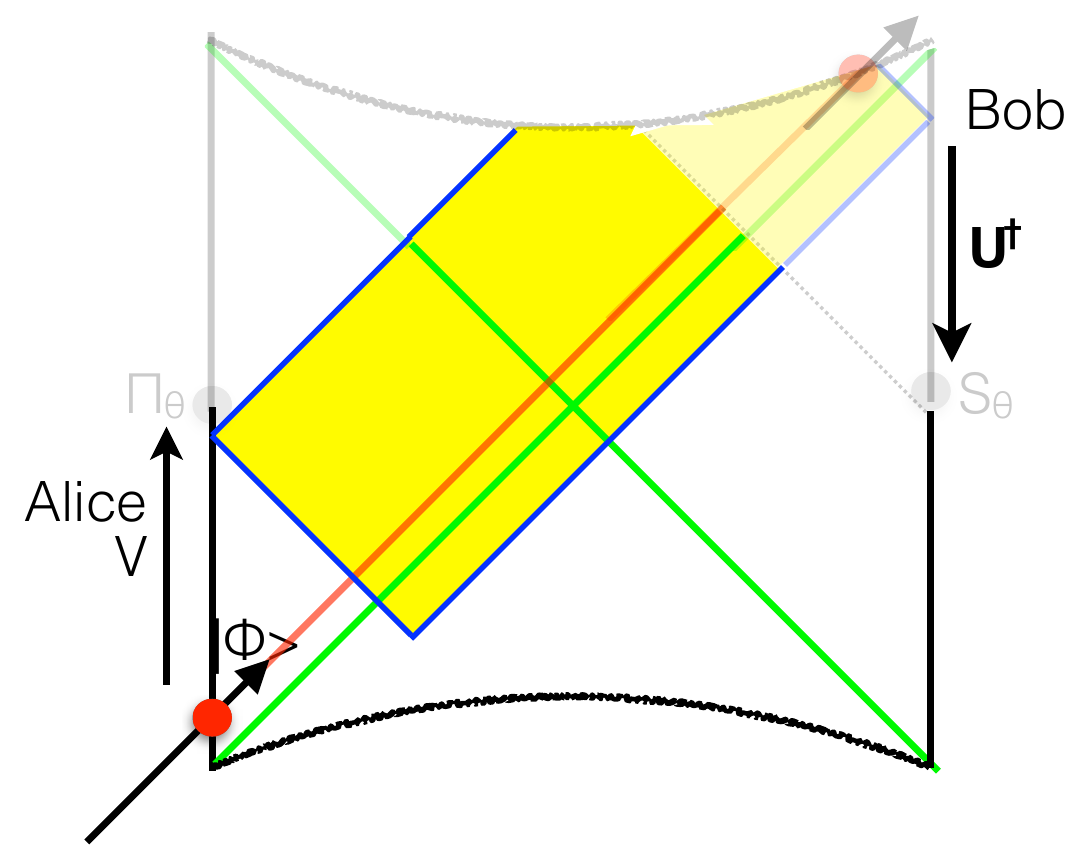}
    \caption{Before the interaction}
    \label{WDWtraversabledecode1}
    \end{center}
  \end{subfigure}
  \hspace{-1cm}
  \begin{subfigure}[b]{0.4\textwidth}
  \begin{center}
    \includegraphics[scale=0.5]{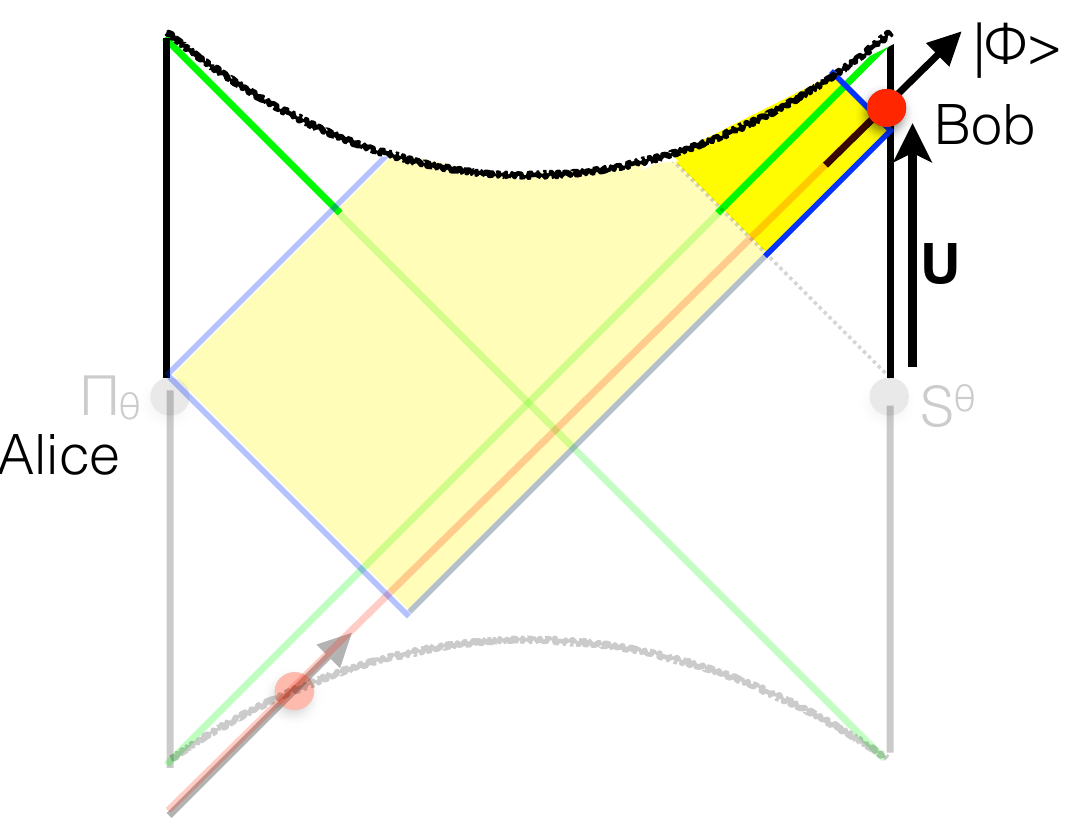}
    \caption{After the interaction}
    \label{WDWtraversabledecode2}
    \end{center}
  \end{subfigure}
  \hspace{-1.8cm}
  \begin{subfigure}[b]{0.5\textwidth}
  \begin{center}
    \includegraphics[scale=0.45]{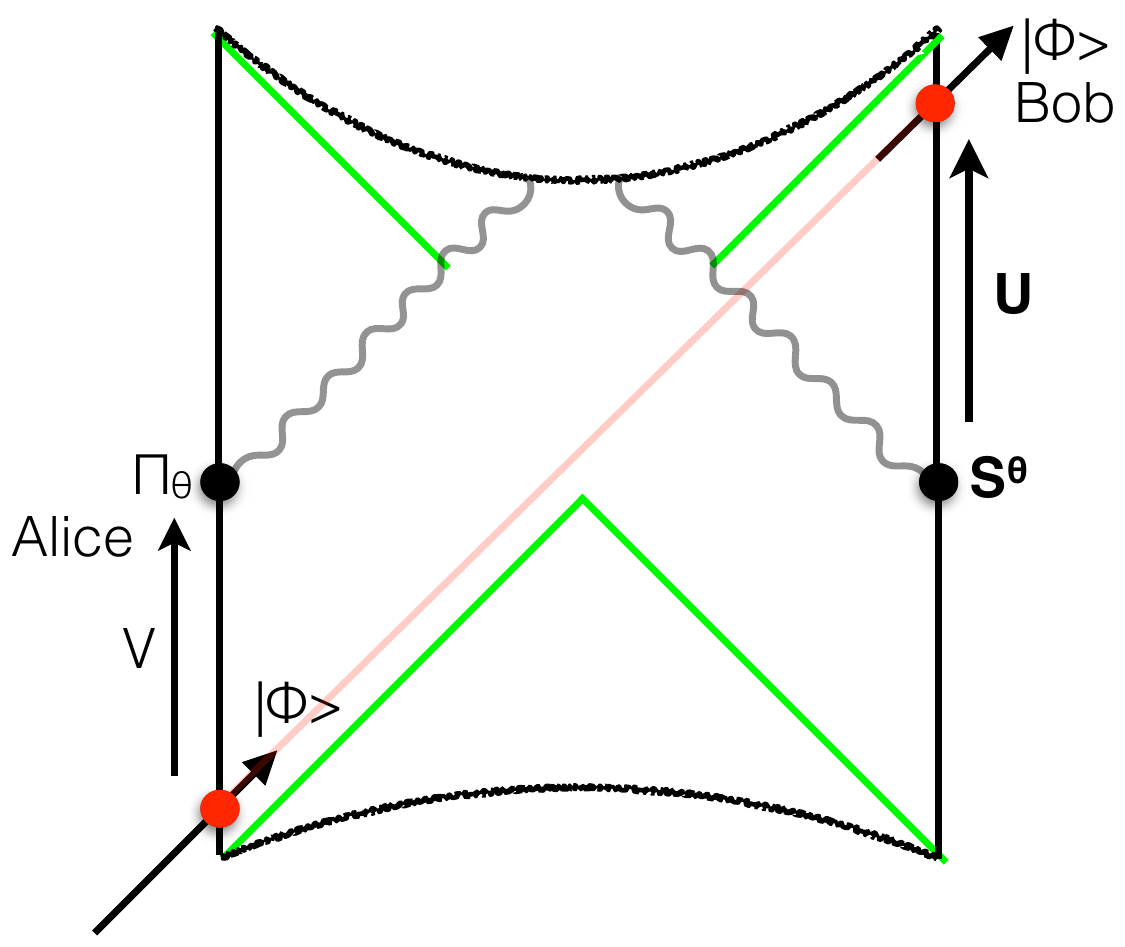}
    \caption{}
    \label{WDWtraversabledecode3}
    \end{center}
  \end{subfigure}
  \hspace{-3cm}
  \caption{Decoding by Bob}
  \label{WDWtraversabledecode}
  \end{center}
  \vspace{-.5cm}
\end{figure}

If we look at the dual geometry of the traversable wormhole (Figure \ref{WDWtraversabledecode}), we see that Bob's decoding operation is $US_{\theta}U^{\dagger}$. (Figure \ref{WDWtraversabledecode1} is the Penrose diagram before the interaction, and we complete it to the future. Figure \ref{WDWtraversabledecode2} is the Penrose diagram after the interaction, and we complete it to the past. Figure \ref{WDWtraversabledecode3}\footnote{In all other figures of the paper, we show the displacement of the message along the shockwave. In this one however, we show a true Penrose diagram. From this we see the shift of the entangling surface and entanglement wedges, which is responsible for the message transfer from Alice to Bob's hands.} pastes these two along the negative energy shells. Combining these, we see that Bob's decoding operation is $US_{\theta}U^{\dagger}$.) This is a precursor. If $U$ lasts for scrambling time and $S_{\theta}$ has complexity of order $1$, the precursor has complexity of order $N$. This is what happens in \cite{Gao:2016bin} \cite{Maldacena:2017axo}. Later, we'll find a protocol, in which $U$ last for half a scrambling time and $S_{\theta}$ has complexity of order $\sim\sqrt{N}$. Both are consistent with our estimate above. 

\section{Complexity of teleportation in a lab}\label{App: labcomplexity}

Say, one has a strongly coupled CFT system, then one can carry out the teleportation in the lab. How complex it is? In order to estimate the complexity of carrying out the protocols in the lab, we impose the following rules. 

\begin{enumerate}

\item 	Complexity accumulated during the natural evolution does not count. 
\item 	Complexity of classical communication is negligible.
\item 	Time $< 0$ is fictitious. Acting at $t<0$ always means acting with a precursor. We have to count the precursor complexity.
\item 	Bob cannot receive the message before Alice sends it.

\end{enumerate}

These rules are not boost invariant, and depend on a particularly chosen slice $t=0$, which we are free to choose in the lab. Let's see how the choice of this particular slice affect the complexity. By varying the choice of the slice we want to minimize the complexity.

\begin{figure}[H]
 \begin{center}
 \hspace{-2.5cm}
  \begin{subfigure}[b]{0.4\textwidth}
  \begin{center}
    \includegraphics[scale=0.45]{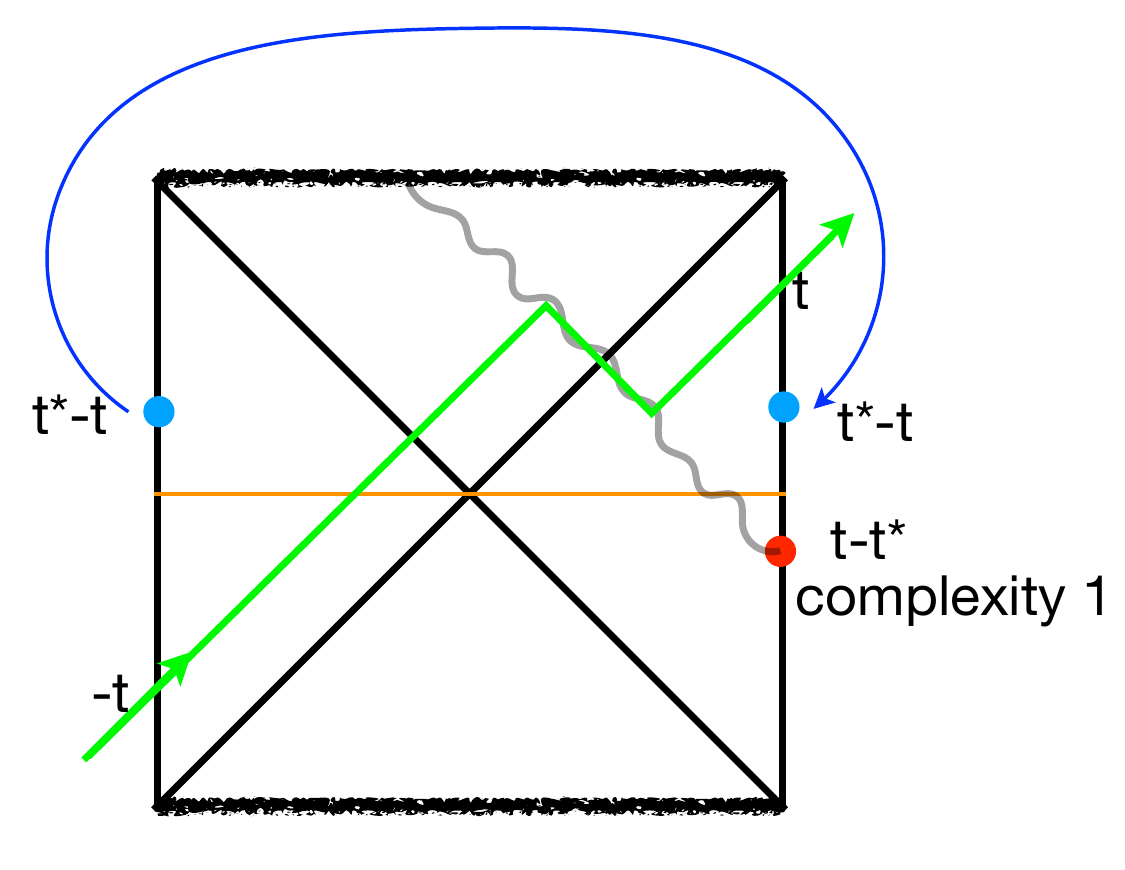}
    \caption{$t>\frac{1}{2}t_*$}
    \label{time1}
    \end{center}
  \end{subfigure}
  \begin{subfigure}[b]{0.4\textwidth}
  \begin{center}
    \includegraphics[scale=0.45]{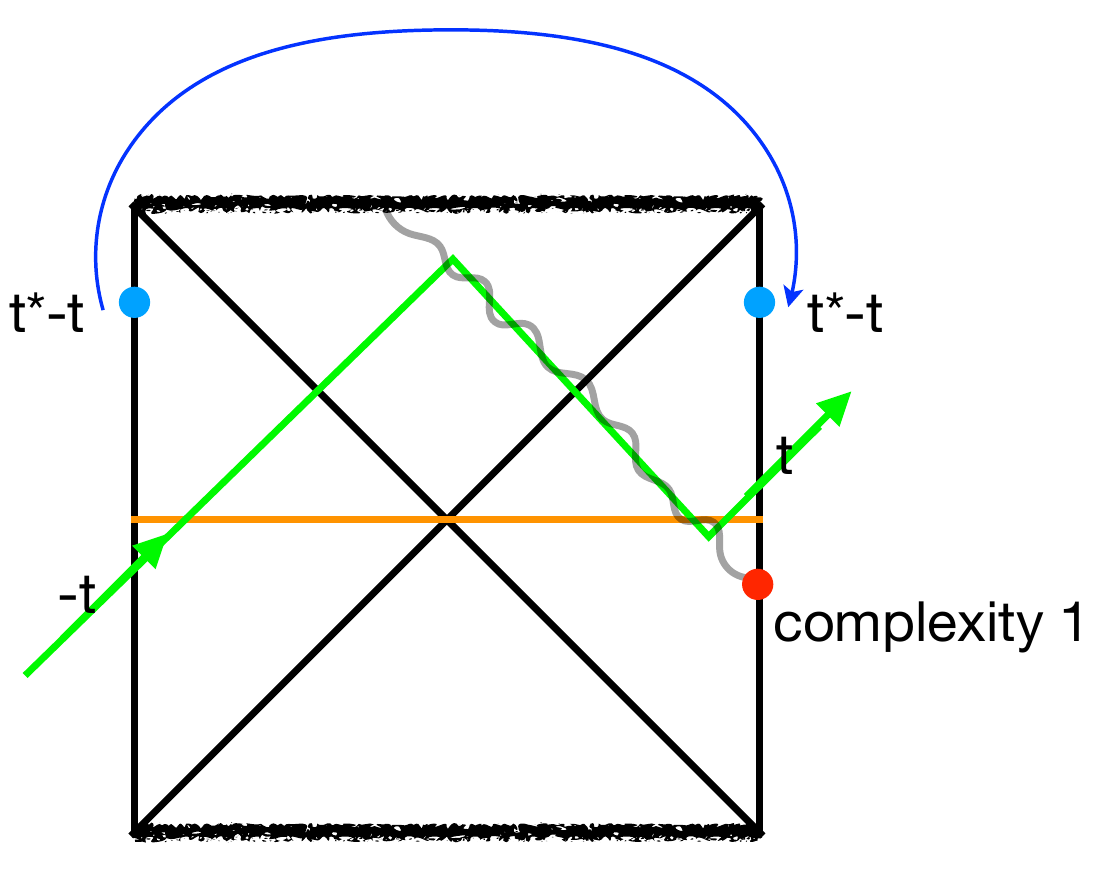}
    \caption{$t<\frac{1}{2}t_*$}
    \label{time2}
    \end{center}
  \end{subfigure}
  \hspace{-1.8cm}
  \caption{}
  \label{timeMSY}
  \end{center}
  \vspace{-.5cm}
\end{figure}

Assume the qubit is sent in by Alice at time $-t$, where $0<t<t_*$. Her encoding complexity is $C_1 = e^t$. Alice sends Bob classical message at $t_*-t$. Bob should get the qubit at time $t$. According rule (4), he can receive Alice' classical message and start decoding no earlier than $t_*-t$. Note that if $t<\frac{1}{2}t_*$, $t_*-t>t$ (Figure \ref{time2}), Bob's decoding precursor will make the qubit come out in his past. If we want to forbid this, we need to have $t>\frac{1}{2}t_*$ (Figure \ref{time1}). \\

The decoding complexity in the protocol in \cite{Gao:2016bin} \cite{Maldacena:2017axo} is
\begin{align*}
C_2 = N\log(1+\frac{1}{N}e^{2(t^*-t)})
\end{align*}

The total complexity (including Alice' encoding and Bob's decoding) will be
\begin{align*}
C =\ &C_1+C_2 =e^t+N\log(1+\frac{1}{N}e^{2(t^*-t)}) = \sqrt N\left[\frac{e^t}{\sqrt N}+\sqrt N\log(1+\frac{N}{e^{2t}})\right]\\
\end{align*}

It takes minimal value when $t = \frac{2}{3}t_*$. The minimal total complexity is $2N^{\frac{2}{3}}$. 

\begin{figure}[H]
 \begin{center}
 \hspace{-2.5cm}
  \begin{subfigure}[b]{0.4\textwidth}
  \begin{center}
    \includegraphics[scale=0.45]{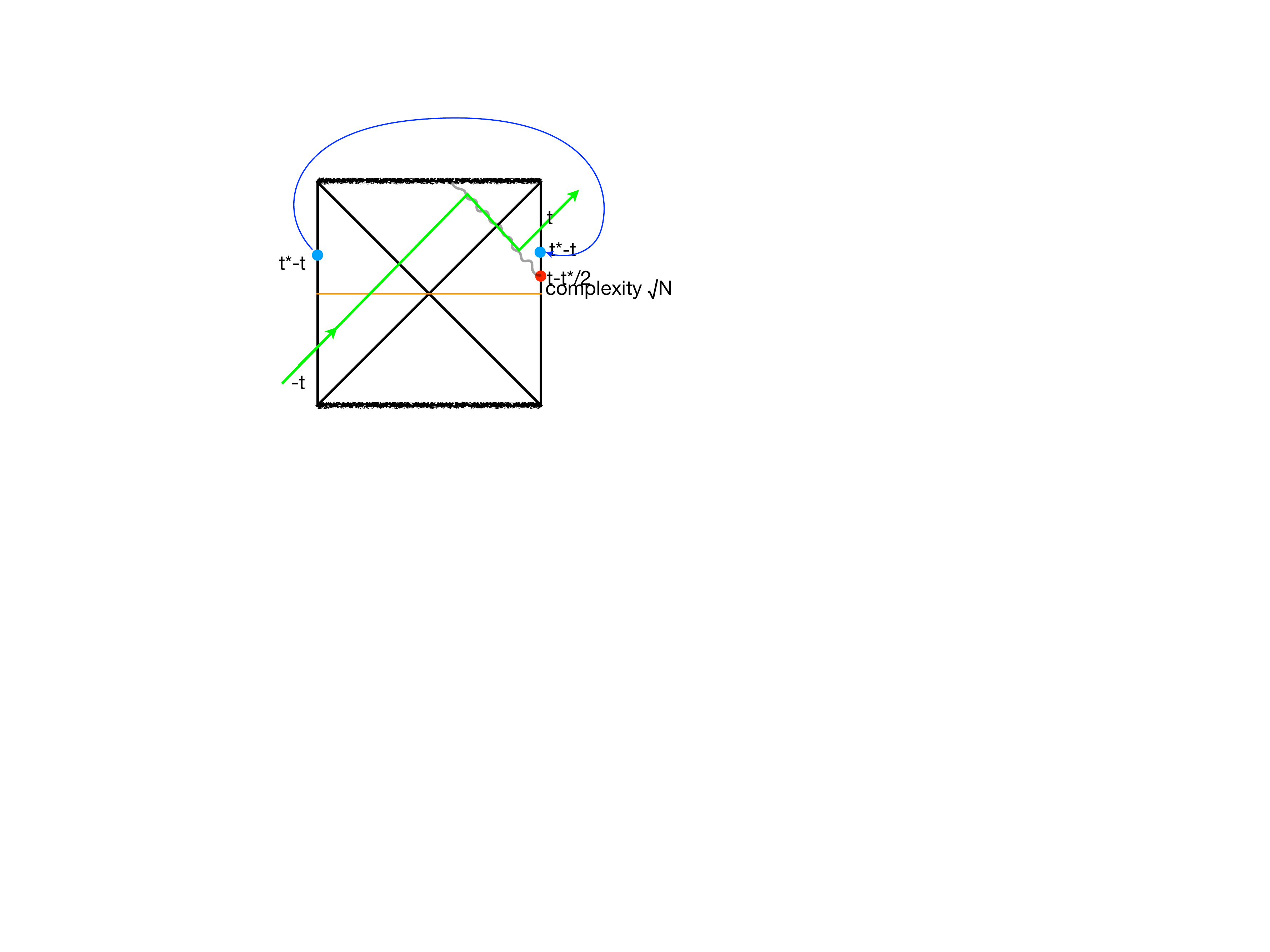}
    \caption{$t<\frac{3}{4}t_*$}
    \label{time3}
    \end{center}
  \end{subfigure}
  \begin{subfigure}[b]{0.4\textwidth}
  \begin{center}
    \includegraphics[scale=0.45]{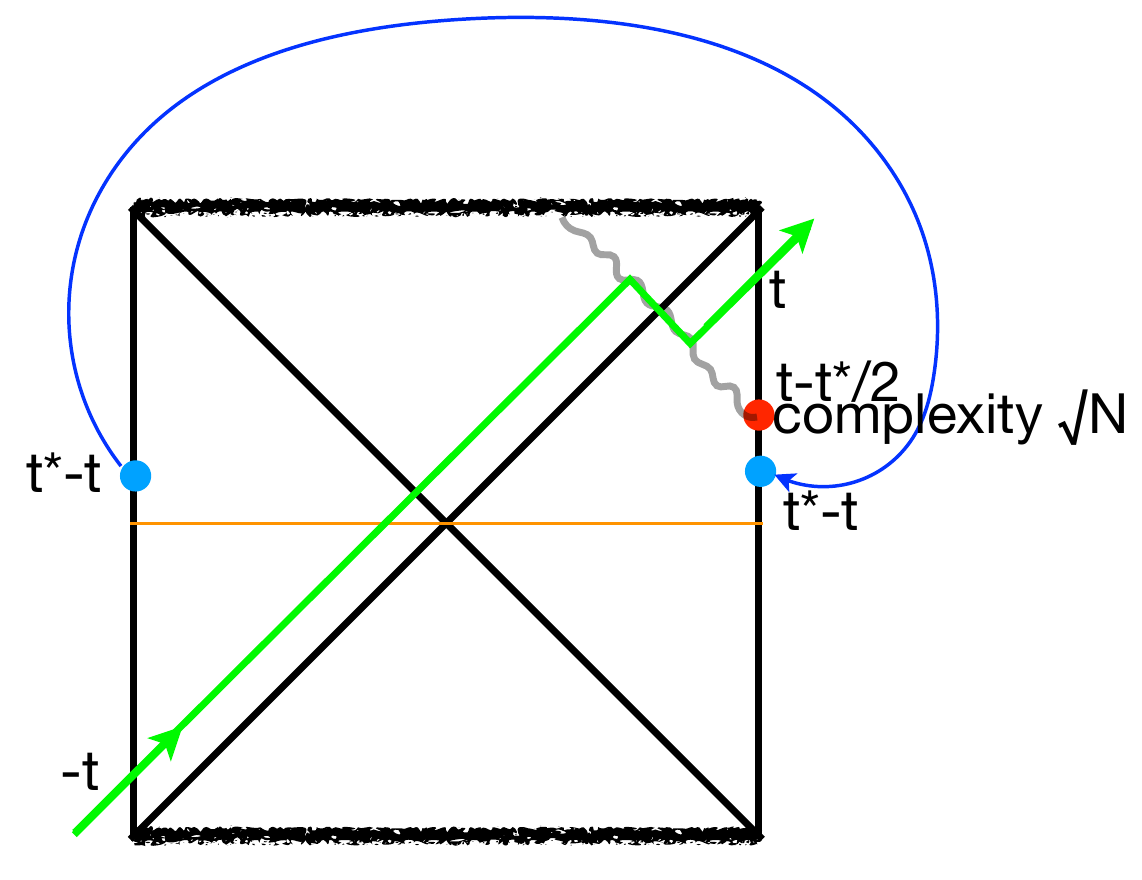}
    \caption{$t>\frac{3}{4}t_*$}
    \label{time4}
    \end{center}
  \end{subfigure}
  \hspace{-1.8cm}
  \caption{}
  \label{timeSZ}
  \end{center}
  \vspace{-.5cm}
\end{figure}

The complexity in our protocol is slightly different: 

Bob's decoding operation has minimal complexity $\sqrt N$ when it is applied at $t-\frac{1}{2}t_*$. On the other hand, Bob's decoding operation can be applied no earlier than $t_*-t$. If $t>\frac{3}{4}t_*$, $t-\frac{1}{2}t_*>t^*-t$, (Figure \ref{time4}), Bob can wait until $t-\frac{1}{2}t_*$ and apply his operator with complexity $\sqrt N$. 

Ignoring forward time evolution, Bob's decoding takes complexity
\begin{align*}
C_2 = 
\begin{cases}
N\log(1+\frac{1}{N}e^{2(t_*-t)}) & t<\frac{3}{4}t_*\\
\sqrt N  & t>\frac{3}{4} t_*
\end{cases}
\end{align*}

The total complexity needed in the lab is
\begin{align*}
C = \begin{cases}
\sqrt N\left(\frac{e^t}{\sqrt N}+\sqrt N\log(1+\frac{N}{e^{2t}})\right) & t<\frac{3}{4}t_*\\
\sqrt N\left(\frac{e^t}{\sqrt N}+1\right) & t>\frac{3}{4}t_*
\end{cases}
\end{align*}

Comparing two protocols, when $t<\frac{4}{3}t_*$, the complexity of two protocols are the same. When $t>\frac{4}{3}t_*$, \cite{Gao:2016bin} \cite{Maldacena:2017axo} has lower complexity. Because Bob's decoding complexity there can keep decreasing until of order $1$, while in our protocol, Bob's decoding complexity cannot decrease below $\sqrt N$. 

We also see that same as in \cite{Gao:2016bin} \cite{Maldacena:2017axo}, the minimal still happens at $t = \frac{2}{3}t_*$, when the minimal complexity is $2 N^{\frac{2}{3}}$.

If someone wants to carry out this teleportation in a lab, he should choose to send the teleportee at time $-\frac{2}{3}t^*$, and the teleporee will emerge on the other side at time $+\frac{2}{3}t_*$. The complexity will be $\sim N^{\frac{2}{3}}$. It's much smaller than the entropy of the CFT system ($\sim N$), but still much larger than the entropy of the teleportee Tom. However, one gets the reward that, by asking Tom about his experience one learns about the interior of a wormhole.

\end{document}